\documentclass[usenatbib]{mn2e}

\usepackage{psfig}

\def\gs{\mathrel{\raise0.35ex\hbox{$\scriptstyle >$}\kern-0.6em\lower0.40ex\hbox{{$\scriptstyle \sim$}}}}
\def\ls{\mathrel{\raise0.35ex\hbox{$\scriptstyle <$}\kern-0.6em\lower0.40ex\hbox{{$\scriptstyle \sim$}}}}
\def\kms{\,\hbox{km}\,\hbox{s}^{-1}}
\def\WoHd{\,\hbox{W$_{o}$}\,\hbox{(H$\delta$)}}

\def\Wm2{\,\hbox{W}\,\hbox{m}^{-2}}
\def\gsim{\mathrel{\raise0.35ex\hbox{$\scriptstyle >$}\kern-0.6em\lower0.40ex\hbox{{$\scriptstyle \sim$}}}}
\def\lsim{\mathrel{\raise0.35ex\hbox{$\scriptstyle <$}\kern-0.6em\lower0.40ex\hbox{{$\scriptstyle \sim$}}}}

\begin{document}

\title[Integral Field Spectroscopy of E+A Galaxies]{From Star-Forming
  Spirals to Passive Spheroids: Integral Field Spectroscopy of E+A
  Galaxies}

\author[Swinbank et al.]
{ \parbox[h]{\textwidth}{ 
A.\,M.\ Swinbank,$^{\! 1,*}$
M.\,L.\ Balogh,$^{\! 2}$
R.\,G.\ Bower,$^{\! 1}$
A.\,I.\ Zabludoff,$^{\! 3}$
J.\,R.\ Lucey,$^{\! 1}$
S.\,L.\ McGee,$^{\! 1}$
C.\,J.\ Miller,$^{\! 4}$
\& R.\,C.\ Nichol$^{\! 5}$
}
\vspace*{6pt}\\
$^1$Institute for Computational Cosmology, Department of Physics and Astronomy, University of Durham, South Road, Durham DH1 3LE, UK\\
$^2$Department of Physics, University of Waterloo, Waterloo, ON, Canada N2L 3G1\\
$^3$Steward Observatory, University of Arizona, 933 North Cherry Avenue, Tucson, AZ 85721-0065, USA\\
$^4$Astronomy Department, University of Michigan, Ann Arbor, MI 48109, USA \\
$^5$Institute for Cosmology and Gravitation, Mercantile House, Hampshire Terrace, University of Portsmouth, Portsmouth, UK PO1 2EG\\
$^*$email: a.m.swinbank@dur.ac.uk\\
}

\maketitle

\begin{abstract} 

We present three dimensional spectroscopy of eleven E+A galaxies at
$z$=0.06--0.12.  These galaxies were selected for their strong
H$\delta$ absorption but weak (or non-existent) [O{\sc
    ii}]$\lambda$3727 and H$\alpha$ emission.  This selection suggests
that a recent burst of star-formation was triggered but subsequently
abruptly ended.  We probe the spatial and spectral properties of both
the young ($\lsim$1\,Gyr) and old ($\gsim$few Gyr) stellar populations.
Using the H$\delta$ equivalent widths we estimate that the burst masses
must have been at least 10\% by mass (M$_{\rm
  burst}\gsim$10$^{10}$\,M$_{\odot}$), which is also consistent with
the star-formation history inferred from the broad-band SEDs.  On
average the A-stars cover $\sim33$\% of the galaxy image, extending
over 2--15\,kpc$^2$, indicating that the characteristic E+A signature
is a property of the galaxy as a whole and not due to a heterogeneous
mixture of populations.  In approximately half of the sample, we find
that the A-stars, nebular emission, and continuum emission are not
co-located, suggesting that the newest stars are forming in a different
place than those that formed $\lsim$1\,Gyr ago, and that recent
star-formation has occurred in regions distinct from the oldest stellar
populations.  At least ten of the galaxies (91\%) have dynamics that
class them as ``fast rotators'' with magnitudes, v/$\sigma$,
$\lambda_R$ and B/T comparable to local, representative ellipticals and
S0's.  We also find a correlation between the spatial extent of the
A-stars and the dynamical state of the galaxy such that the fastest
rotators tend to have the most compact A-star populations, providing
new constraints on models that aim to explain the transformation of
later type galaxies into early types.  Finally, we show that there are
no obvious differences between the line extents and kinematics of E+A
galaxies detected in the radio (AGN) compared to non-radio sources,
suggesting that AGN feedback does not play a dramatic role in defining
their properties, and/or that its effects are short.

\end{abstract}


\section{Introduction}

Galaxies with strong Balmer absorption lines in their spectra, but weak
nebular emission (such as [O{\sc ii}]$\lambda$3727\AA\,), represent a
short-lived but potentially important phase in galaxy evolution
\citep[e.g.\ ][]{Tran03,Tran04}.  These ``E+A'' galaxies have strong
absorption lines (such as H$\delta$), representing a stellar population
dominated by A-stars, which are either absent or overwhelmed by the
much brighter OB stars in most galaxies.  These signatures suggest that
the star-formation within the galaxy abruptly ended $\lsim$1\,Gyr ago,
possibly following a starburst phase.  It is likely that a variety of
physical mechanisms lead to such a stellar population
(e.g. \citealt{Dressler82,Couch87,Zabludoff96,Poggianti99,Balogh99}),
but most invoke a major transformation from one galaxy type to another,
possibly representing an evolutionary link between gas-rich,
star-forming galaxies and quiescent spheroids.  Although such galaxies
are very rare in the local Universe, their short lifetime means they
could potentially represent an important phase in the evolution of most
normal galaxies \citep[e.g.][]{Yan09,Zabludoff96}.

In order to trace the route by which star-forming galaxies evolve into
quiescent systems, there are several issues which must be addressed.
First, from where in the galaxy does the unusual spectrum originate?
Is the strong H$\delta$ a result of galaxy wide A-star populations, or
is it due to a heterogeneous mixture of populations, such as a nuclear
starburst with a very high equivalent width ($>$10\AA) on top of an
otherwise normal galaxy?  Second, what star formation history leads to
the spectral characteristics of these galaxies?  Is the E+A phase
related to a significant enhancement of the star formation, or simply
the rapid decline of star formation in normal, gas-rich
galaxies.  Finally, what triggers the recent change in star formation
history?  

Several physical mechanisms have been suggested for the triggering
(and subsequent rapid end) of starbursts in field E+As, including
major mergers \citep{Mihos96} and unequal mass mergers and tidal
interactions \citep{Bekki05,Pracy05}.  Recently, large-scale energetic
outflows from AGN have been invoked as a route to explain many of the
properties of local massive galaxies.  If post-starbursts are
late-time mergers they may have passed through their quasar phase and,
as a result, the AGN may provide sufficient energy to inhibit and
terminate star-formation \citep{Bower06}. Indeed, post-starburst
galaxies have been suggested as ideal laboratories for testing AGN
feedback models \citep{Tremonti07}, and some observations indicate
that such a link exists \citep{Georgakakis98,Yan06}. Finally, in dense
environments (such as clusters) galaxy harassment and/or interaction
with the hot intra-cluster gas is also likely a driving factor
\citep{Dressler83,Pracy10}.

While high resolution imaging has demonstrated that most bright, nearby
E+A galaxies are spheroidal, often with signs of interaction
\citep{Yang04,Balogh04}, their dynamics provide better constraints on
their likely origins.  \citet{Norton01} obtained longslit spectra of
twenty E+A galaxies and used the ratio of the rotational velocity to
line-width to conclude that most of the galaxies are in the process of
transforming from rotationally-supported, gas-rich galaxies to
pressure-supported, gas-poor galaxies.  However, longslit spectroscopy
mixes spatial and spectral resolution, and a better understanding of
the dynamics can be obtained from integral field spectroscopy, where
the spatial and spectral information can be cleanly decoupled.  This
technique also allows the location, spatial extent and separation of
any gas, old (K) stars and (the short-lived) massive A-stars to be
compared.  

An example of such measurements were made by \citet{Swinbank05a} using
the Gemini/GMOS-South IFU to map the properties of a H$\delta$ strong
galaxy (SDSS\,J1013+0116) selected from the Sloan Digital Sky Survey
(SDSS).  In this galaxy, the velocity field and spatial extent of the
distribution of H$\delta$ equivalent widths suggest that the
post-starburst phase arose from a merger between a gas-rich spiral and
a gas-poor, passive secondary.  Crucially, within SDSS\,J1013+0116, the
A-stars are distributed throughout the galaxy, arguing that the
starburst occurs in the disk of the galaxy during the interaction.  The
studies of \citet{Goto08a} and \citet{Pracy09} increased the number of
E+A galaxies with resolved spectroscopy to $\sim$10, and showed that
wide-spread A-star populations are a common feature, with A-star
populations extended on $>>$\,kpc scales.  \citet{Pracy09} also showed
that the majority of their E+A galaxies have significant angular
momentum per unit mass (so-called 'fast-rotators'), casting doubt on
equal-mass, major mergers on the mechanism which results in E+A's since
these should leave dispersion-dominated kinematics.

In order to gain a better understanding of the interaction between
star-formation and gas dynamics within the ISM of E+A galaxies, we have
obtained three dimensional spectroscopy of eleven galaxies, selected
from the SDSS in the redshift range $0.07<z<0.12$.  The three
dimensional data are used to map the distribution and kinematics of
A-stars.  We investigate the spatial distribution of
residual star formation through the [O{\sc ii}]$\lambda$3727 emission
and the distribution of the young A-stars through the much stronger
H$\delta$ absorption line.  In \S2 we present the data reduction and
analysis.  The results are presented in \S3.  Finally we summarize our
results and present the implications in \S4.  We adopt a cosmology with
$\Omega_{\rm m}$=0.27, $\Lambda=1-\Omega_{\rm m}$ and a Hubble
constant of 72$\kms$Mpc$^{-1}$. In this cosmology, at $z=0.09$ (the
median redshift of our sample), 1\,kpc subtends 0.6$''$ on the sky.

\begin{table}
{\scriptsize
\begin{center}
{\centerline{\sc Table 1: Target Information}}
\smallskip
\begin{tabular}{lcccccccccccc}
\hline
\noalign{\smallskip}
Target           & $\alpha_{J2000}$ & $\delta_{J2000}$       & redshift & Exptime \\
                 & $h\,m\,s$        & $^{\circ}$\,$'$\,$''$ &            & $(ks)$  \\
\hline                                                                  
J0835+4239       & 08\,35\,21.23    & +42\,39\,36.0        & 0.0920   & 10.8 \\
J0906+5221$^{*}$ & 09\,06\,19.93    & +52\,21\,50.0        & 0.0989   & 10.8 \\
J0938+0001       & 09\,38\,42.91    & +00\,01\,49.0        & 0.0914   & 18.0 \\
J0948+0230       & 09\,48\,18.68    & +02\,30\,04.1        & 0.0602   & 16.2 \\
J1013+0116       & 10\,13\,45.39    & +01\,16\,13.6        & 0.1055   & 10.8 \\
J1242+0237$^{*}$ & 12\,42\,52.96    & +02\,37\,00.9        & 0.0846   & 10.4 \\
J1348+0204       & 13\,48\,02.21    & +02\,04\,05.6        & 0.0678   & 10.8 \\
J1627+4800       & 16\,27\,55.98    & +48\,00\,51.1        & 0.1254   & 13.0 \\
J1642+4153       & 16\,42\,55.22    & +41\,53\,35.5        & 0.0722   & 11.7 \\
J1715+5822       & 17\,15\,46.24    & +58\,22\,55.4        & 0.1271   & 15.6 \\
J2307+1525       & 23\,07\,43.40    & +15\,25\,59.3        & 0.0699   & 14.4 \\
\hline\hline                                                                   
\label{table:obs_log}                                                          
\end{tabular}
\vspace{-0.5cm}
\caption{Co-ordinates, redshifts and exposure times of the galaxies
  in our sample.  The two E+A galaxies labeled with a $^{*}$
  (J0906+5221 and J1242+0237) are both in cluster environments.  Note
  that J2013+0116 was selected to have weak emission lines, so is an
  e(a), rather than E+A, galaxy.}
\end{center}
}
\end{table}

\section{Observations and Data Reduction}

\begin{figure}
\centerline{
  \psfig{figure=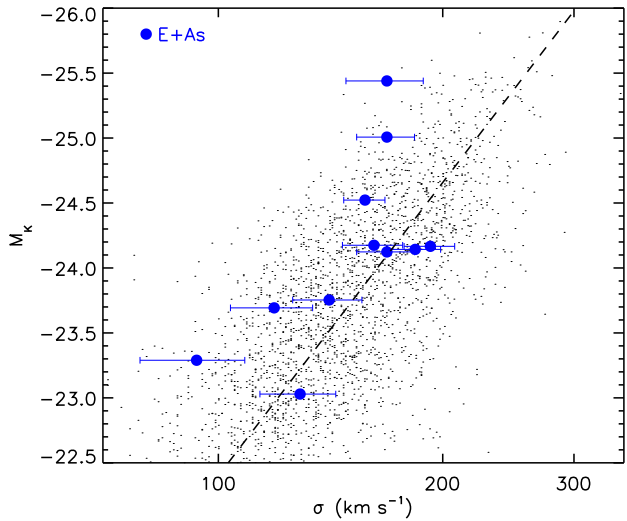,width=3.5in,angle=90}}
\caption{The $K$-band velocity--line-width relation of the E+A galaxies
  in our sample compared to SDSS early type galaxies from
  \citet{Bernardi05}, with  $K$-band magnitudes taken from 2\,MASS.
  This figure shows that the E+A galaxies in our study span
  $\sim$80--180\,km\,s$^{-1}$ and $\gsim$2\,mags in $K$.  Our study
  focuses on the brighter E+A galaxies at a fixed velocity dispersion, 
  due to the requirement of high signal-to-noise ratios to carry out this analysis. }
\label{fig:FP}
\end{figure}

\begin{figure}
\centerline{
  \psfig{figure=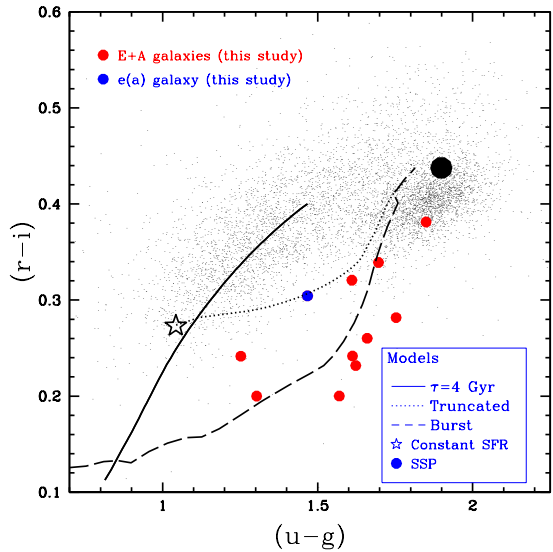,width=3in,angle=0}}
\caption{The observed $(u-g)$ versus $(r-i)$ distribution (corrected
  for Galactic reddening), of galaxies from the SDSS (model magnitudes).
  Small points represent a random subset of galaxies from DR5 with $16<r<17$ in this
  redshift range; the bimodal nature of the population is evident.
  Filled, coloured circles represent our GMOS IFU targets; the red
  points are the E+A galaxies and the blue point is the single e(a)
  galaxy from \citet{Swinbank05a}.  The lines are the predictions of
  \citet{Bruzual03} models, similar to those used in \citet{Balogh04}.
  All assume a Salpeter IMF and represent observed colours at $z=0.09$.
  The filled, black circle represents an old population, resulting from a
  single burst of star formation 13.7 Gyr ago, while the star indicates
  the colour of a galaxy that has undergone constant star formation for
  the same amount of time, and includes internal extinction of
  $\tau_v=1$ mag.  The solid line represents the track of an
  exponentially--declining star formation rate, with $\tau=4$Gyr, and
  $\tau_v=1$mag extinction.  The dotted line is a "truncated" model,
  which shows the colour evolution over a 2\,Gyr period following
  truncation of star formation in the model represented by the open
  star.  Finally, the long--dashed line represents a model in which a
  10 per cent (by mass), instantaneous burst is superposed upon the old
  stellar population (filled circle); the evolution from blue to red is
  followed for 2 Gyr following the burst.  This latter model (which
  includes $\tau_v=1$ mag extinction) provides a reasonable match to
  the colours of our E+A galaxies, as shown in \citet{Balogh04}.}
\label{fig:ugri}
\end{figure}

\subsection{Sample Selection}

Our sample is selected from the catalog of \citet{Balogh04}, which is
based on $K$-band imaging of 222 nearby H$\delta$-strong galaxies from
\citet{Goto03}.  This catalog contains all galaxies for which the rest
frame equivalent width of the H$\delta$ absorption line is $\WoHd
>$4\AA\ and W$_{\rm o}$([O{\sc ii}])$<$10\AA\ (with $>$2$\sigma$
confidence), as measured from the spectrum using a Gaussian profile
fitting technique.  Of these, we select galaxies which also have weak
or absent H$\alpha$ emission, $W_{o}$(H$\alpha<$5\AA).  This helps to
exclude galaxies with weak or dusty star formation, as well as
contribution from an active galactic nucleus \citep{Yan04}.

Ten E+A galaxies from this sample were selected for follow-up IFU
observations.  These galaxies were selected to span different
morphological types and environments.  The former was quantified by
fractional bulge luminosities (B/T) determined from the bulge-to-disk
decompositions of \citet{Balogh04}, while the latter was obtained by
cross-correlating with the C4 cluster catalogue \citep{Miller05}.  In
practice, given our high mass cut, most of the candidates were
bulge--dominated, as shown in Figure~5 of \citet{Balogh04}; we only
found one clearly disk--dominated galaxy (J1048+0230, with B/T=0.29).
Furthermore, the vast majority of the E+A galaxies in \citet{Balogh04}
are not associated with clusters; we found only two galaxies satisfying
our other selection criteria that are near clusters.  In addition, we
include J1013+0116 from \citet{Swinbank05a}, to bring the final sample
to eleven galaxies, as shown in Table~1.  J1013+0116 was chosen for our
pilot programme, for which we required reasonably strong line emission
to ensure reliable dynamical measurements could be made.  Thus, it has
the strongest [O{\sc ii}] emission in our sample (W$_{o}$[O{\sc
    ii}]=5.2$\pm$0.3\AA\,), and is more properly classified as an e(a)
galaxy \citep{Poggianti99}.  This is therefore a fairly heterogeneous
sample, by design.

In Fig.~\ref{fig:FP} we show where these galaxies lie in the
luminosity-line-width plane of the Faber-Jackson relation
\citep{Faber76}.  We measure the velocity dispersions from the SDSS
spectra, using the cross correlation technique described in
\S~\ref{sec:gmos}.  We use the line width of the Mg{\sc i} triplet as a
proxy for the stellar kinematics of an old stellar population.
The E+A galaxies in our sample span a range of velocity dispersions
from $\sigma\sim$80--180$\kms$, and nearly two magnitudes in the
$K$-band.  For comparison we also plot the early-type galaxies selected
from SDSS \citep{Bernardi05}, cross matched to 2\,MASS.  To ensure a
fair comparison with the E+A galaxies, we limit the redshift range of
the comparison sample to $z<0.1$, and apply no k-corrections to either
sample.  As Fig.~\ref{fig:FP} shows, the E+A galaxies in our sample are
systematically brighter in the $K$-band for a fixed velocity dispersion
($\Delta$M$_{\rm K}$=0.6$\pm$0.2).  A comparable offset in $R$-band
($\Delta$m$_{\rm R}\sim$0.6 mags) was also noted by \citet{Norton01}
who used longslit spectroscopy of twenty E+A galaxies from the sample
described by \citet{Zabludoff96}.

The $(u-g)$ versus $(r-i)$ colours of the galaxies in our sample are
shown in Fig.~\ref{fig:ugri}.  We compare these to the galaxies from
SDSS DR5 with redshift range $0.05<z<0.1$ and magnitude range
$16<r<17$.  We also compute three evolutionary colour tracks: i) an
exponentially--declining star formation rate, with $\tau=4$\,Gyr, and
$\tau_v=1$\,mag extinction; ii) a "truncated" model which shows the
colour evolution over a 2\,Gyr period following truncation of star
formation in the model; iii) a model in which a 10 per cent (by mass),
instantaneous burst is superposed upon the old stellar population.  As
Fig.~\ref{fig:ugri} shows, this latter model provides the best match to
the colours of our E+A galaxies, as shown in \citet{Balogh04}.

Finally, we note that of ten galaxies in our sample that are covered
as part of the VLA Faint Images of the Radio Sky at Twenty-Centimeters
(FIRST) survey, two are detected at $>$5\,sigma (J0948+0230 and
J1013+0116) and two more are detected at 3--5$\sigma$ (0835+4239 and
J1642+4153).  These detections have 1.4\,GHz fluxes in the range
0.5--2.2\,mJy, corresponding to luminosities of
L$_{1.4}$=8--15$\times$10$^{21}$\,W\,Hz$^{-1}$.  We return to this in
\S4.

\begin{figure*}
\centerline{
  \psfig{figure=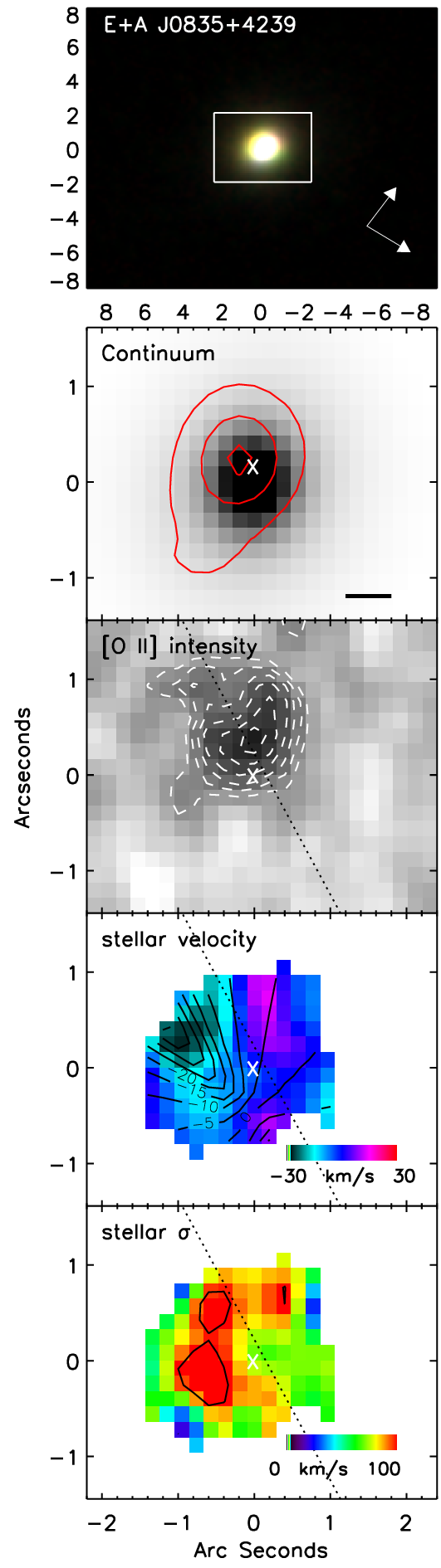,width=1.3in,angle=0}
  \psfig{figure=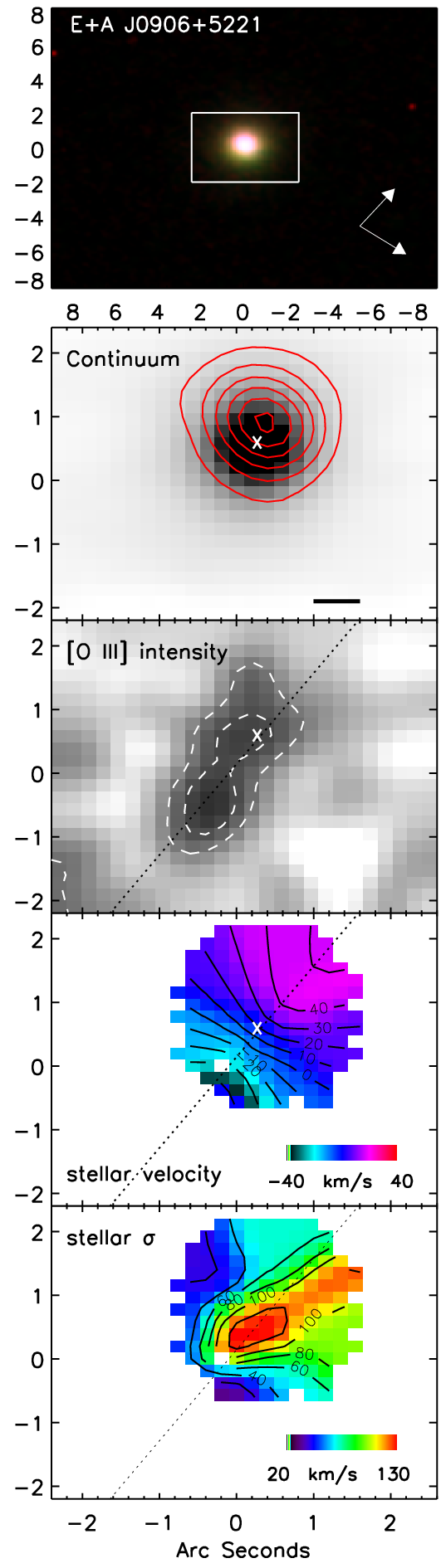,width=1.3in,angle=0}
  \psfig{figure=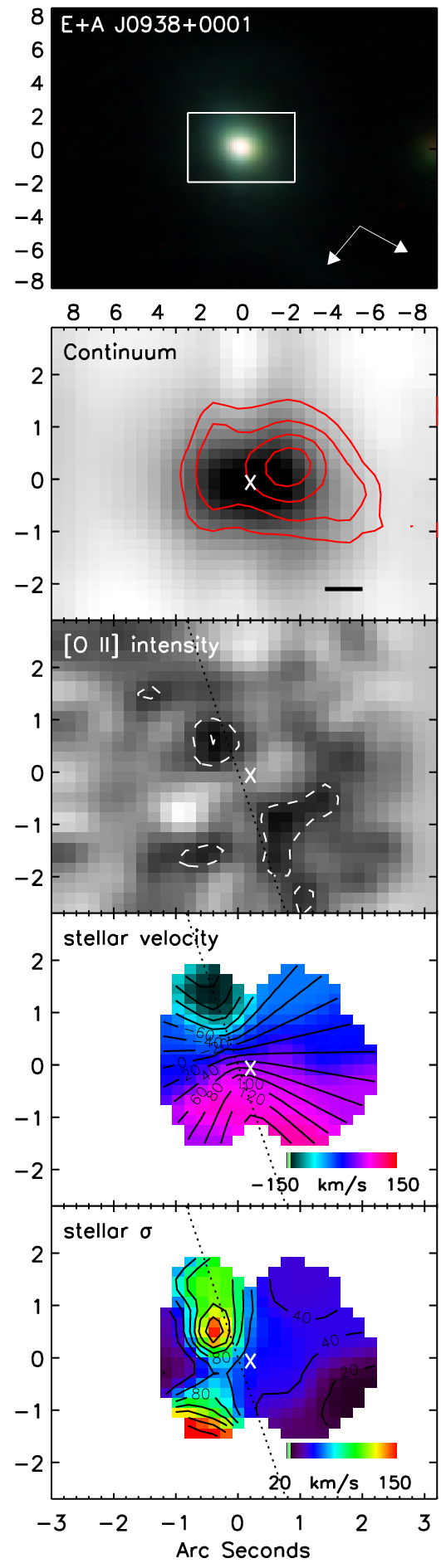,width=1.3in,angle=0}
  \psfig{figure=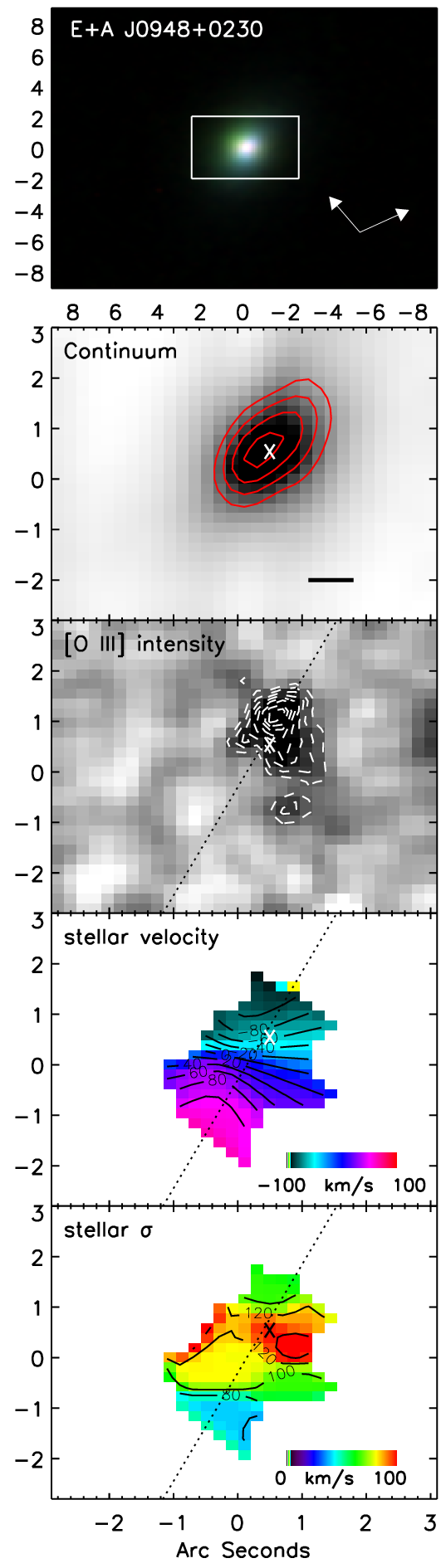,width=1.3in,angle=0}
  \psfig{figure=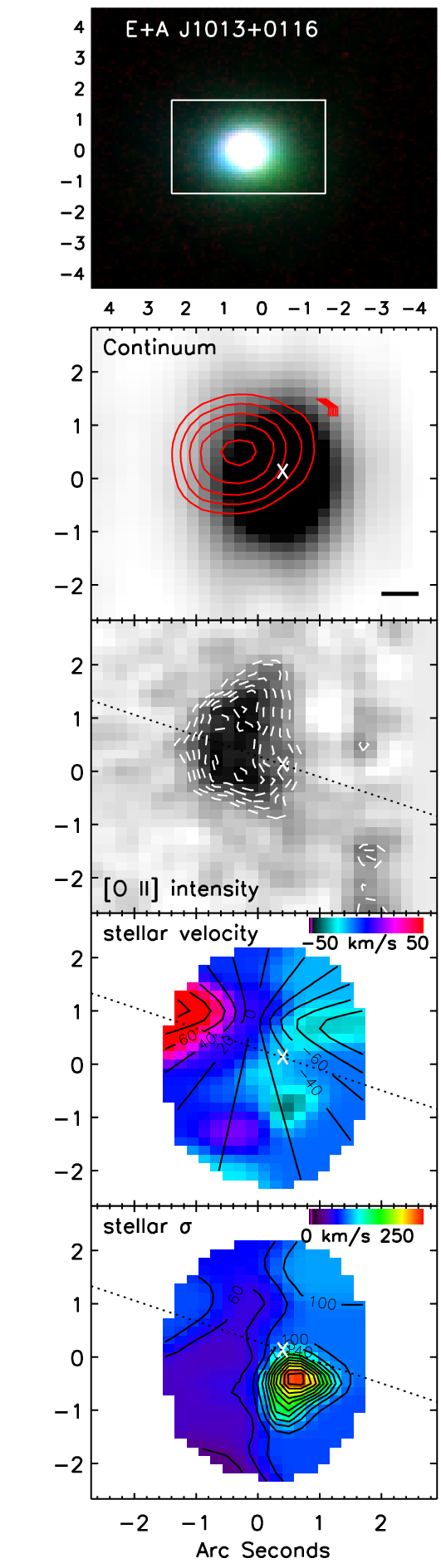,width=1.3in,angle=0}
  \psfig{figure=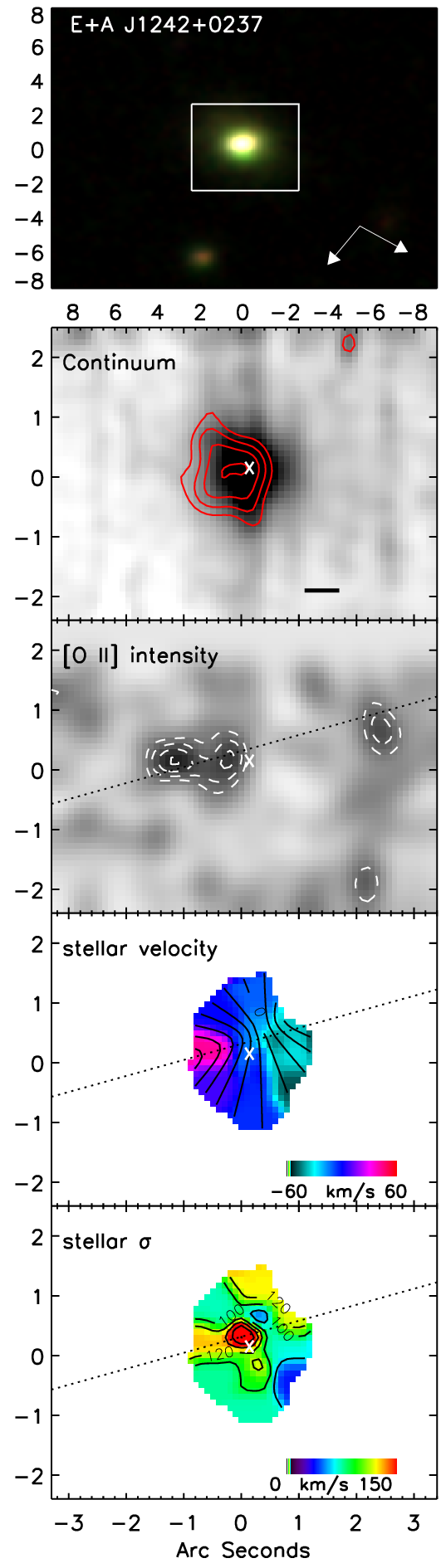,width=1.3in,angle=0}}
\label{fig:2Dmaps}
\caption{Results from the IFU observations of the E+A galaxies in our
  sample.  From top to bottom: 1. True colour $BIK$-band images of
  each of the galaxies constructed from Gemini/GMOS imaging (blue),
  SDSS $i$-band (green) and UKIRT/UFTI $K$-band imaging (red).  In
  each panel we overlay the GMOS IFU field of view.  The arrows denote
  the directions of North and East.  2. Continuum images from the
  continuum above the 4000\AA\ break, derived by collapsing the
  datacube between 4350--4750\AA\ in the rest-frame, with dark grey
  indicating high intensity.  The contours trace the continuum from
  the rest-frame 3650-3850\AA\ continuum emission (which should have a
  stronger contribution from young stars than the continuum emission
  above the 4000\AA\ break).  The solid bar in each panel denotes the
  seeing disk for the observations. 3.  Gas phase emission line
  intensity (derived from either the [O{\sc ii}]$\lambda3727$ or
  [O{\sc iii}]$\lambda$5007, as indicated).  The contours start at
  3$\sigma$ and are incremented by 1$\sigma$.  4.  Two dimensional
  stellar velocity fields of the galaxies, measured from the H$\delta$
  line.  5. Two dimensional stellar velocity dispersion maps.  The
  dotted line marks the major kinematic axis in the velocity fields,
  where it can be clearly identified.}
\label{fig:2Dmaps}
\end{figure*}
\begin{figure*}
\centerline{
  \psfig{figure=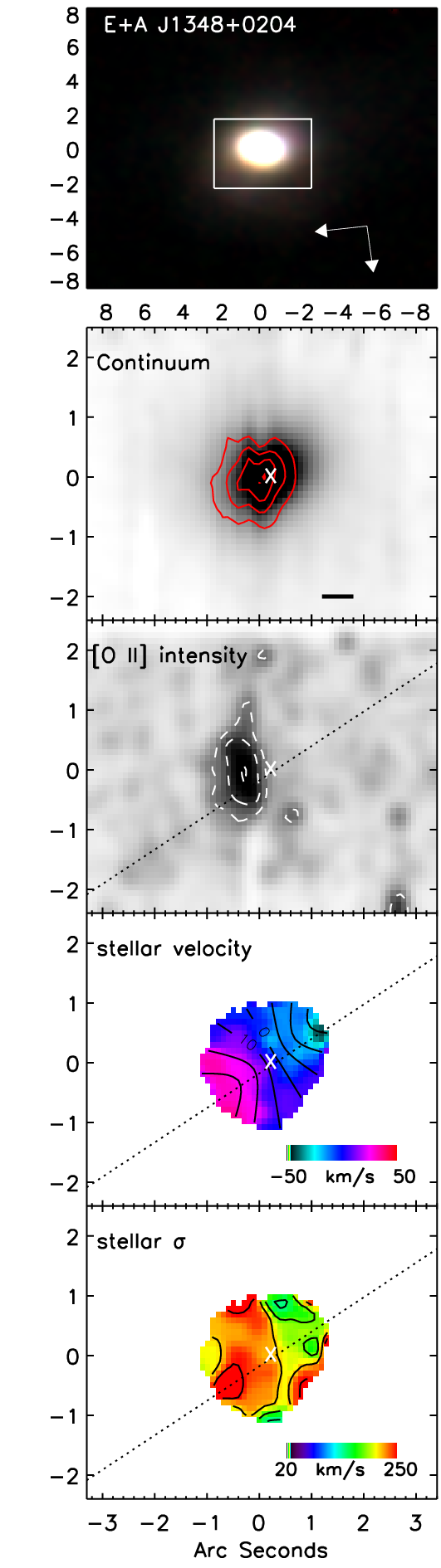,width=1.4in,angle=0}
  \psfig{figure=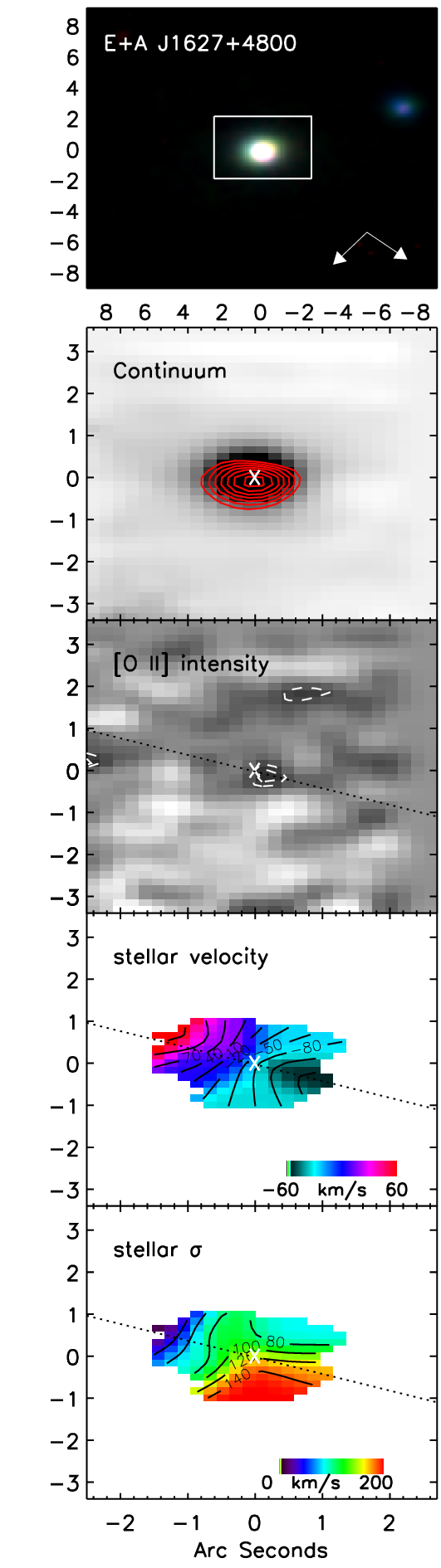,width=1.4in,angle=0}
  \psfig{figure=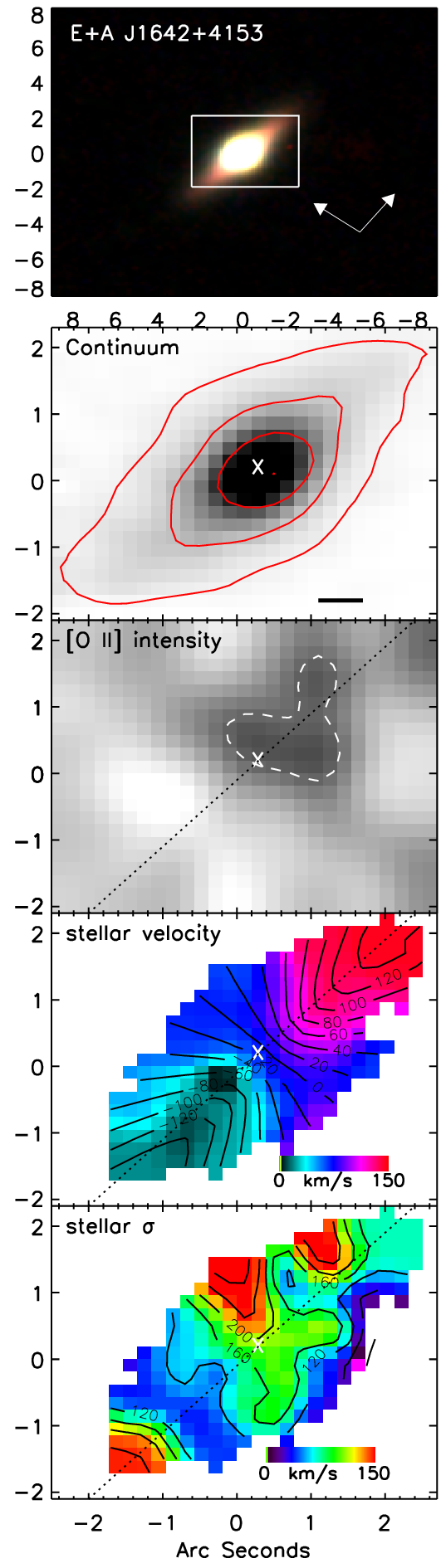,width=1.4in,angle=0}
  \psfig{figure=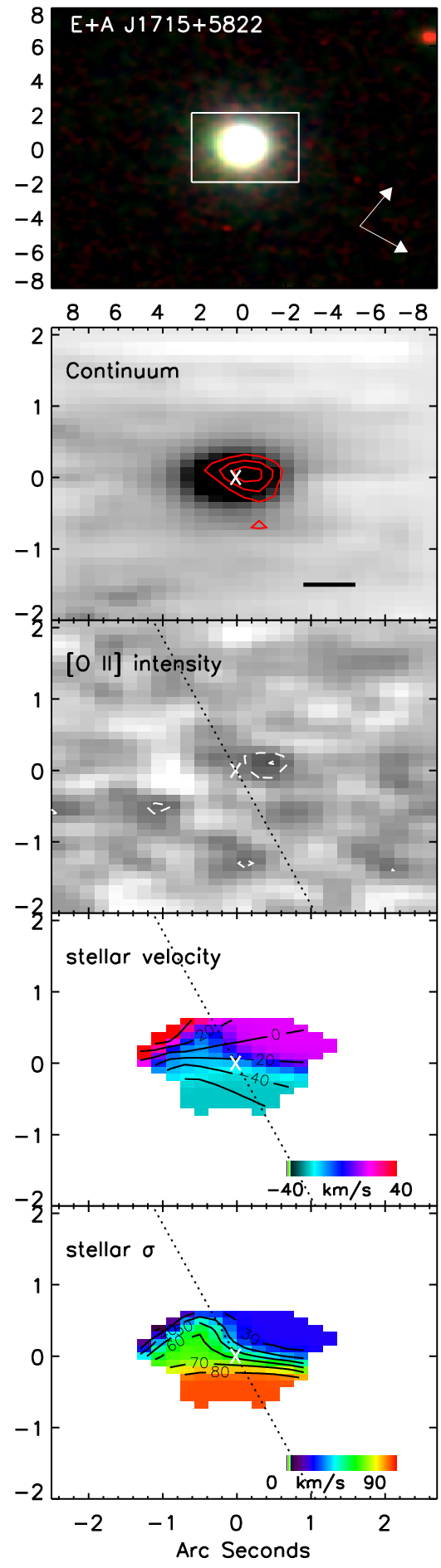,width=1.4in,angle=0}
  \psfig{figure=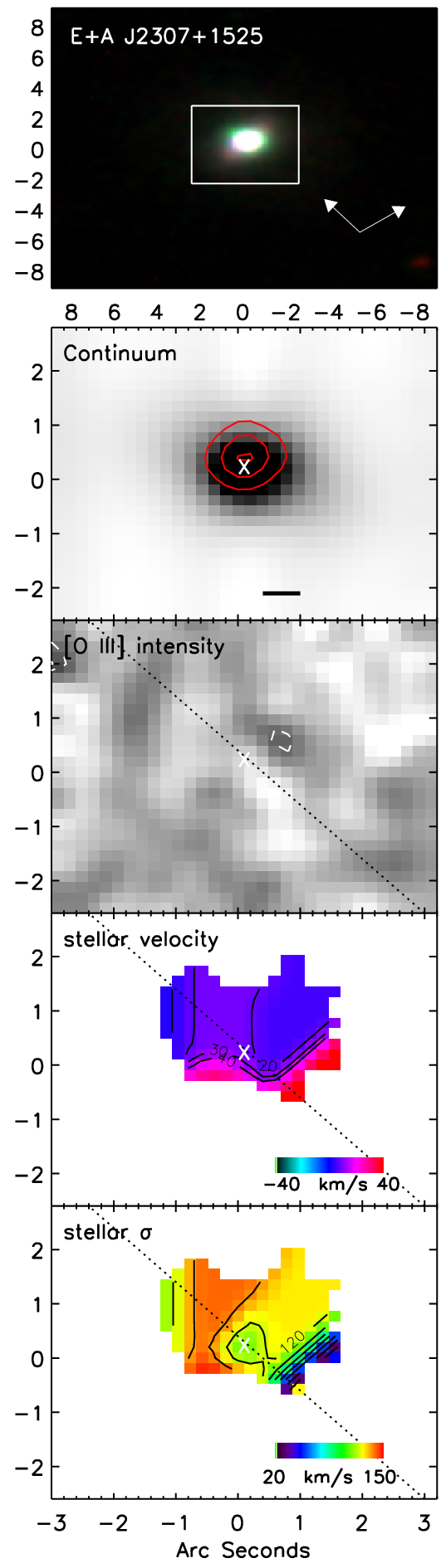,width=1.4in,angle=0}}
\label{fig:2Dmaps_b}
\end{figure*}

\begin{figure*}
\centerline{
  \psfig{figure=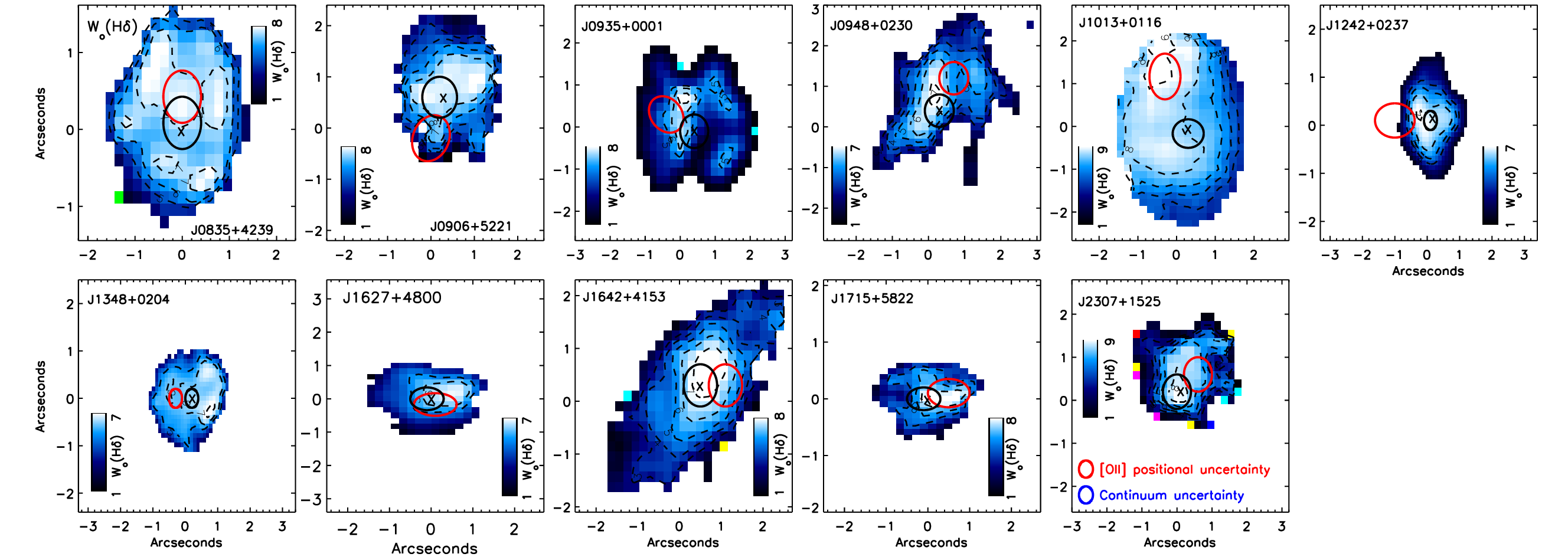,width=8in,angle=0}}
\vspace{0.25cm}
\caption{The distribution of rest-frame H$\delta$ equivalent widths
  across each of the galaxies in our sample.  The contours are
  separated by $\Delta \WoHd$=1\AA.  The 'X' denotes the center of the
  galaxy as defined from the center of the continuum above the
  4000\AA\ break, while the ellipses denote the center and 1$\sigma$
  uncertainty in the nebular emission centroid (red) and continuum
  emission centroid (black).  This figure highlights that six E+As in
  our sample have H$\delta$ absorption, nebular emission and continuum
  emission which are spatially co-located on $\sim$\,kpc scales
  (J0948,J1642,J1627,J1715,J2307), whilst three systems have
  significant offsets between nebular emission and H$\delta$
  (J0906,J1242,J1348).  Two galaxies have nebular emission and
  H$\delta$ in agreement, but with very extended H$\delta$ and nebular
  emission which is offset from the strongest continuum (J0835,J1013),
  suggesting that the A and OB stars formed in a different location
  than the older stars.  At least in some cases, the new OB stars are
  either displaced from the older A-stars.  This reconciles the fact
  that both emission and strong absorption can be seen in the same,
  integrated galaxy spectrum -- the two populations do not generally
  arise from exactly the same place.  }
\label{fig:centers}
\end{figure*}

\subsection {GMOS Spectroscopic Imaging}
\label{sec:gmos}

Spectro-imaging observations of ten new E+A galaxies in our sample were
taken with the GMOS-North and GMOS-South IFU between 2005 and 2007
\citep{AllingtonSmith02}.  Seven targets were observed in `stare' mode
using one-slit mode which results in a field of view of
5$''\times$7$''$, while four targets were observed with nod-and-shuffle
in two-slit mode (resulting in a field of 5$''\times$5$''$).  For
observations taken with nod-and-shuffle we chopped away from the target
by $30''$ every 30 seconds (see \citealt{Swinbank05a} for a detailed
discussion of the observing procedure).  The exposure times for each
target were typically 11-18\,ks, with longer exposures taken on fainter
targets (Table~1).  All observations were carried out in dark time,
typically with 0.6$''$ seeing in $V$-band and in photometric
conditions.  The corresponding physical resolution is $\sim$1.0\,kpc at
the median redshift of our sample.  We used the $B$-band filter in
conjunction with the B600 or B1200 grating, which results in a spectral
resolution of $\lambda/\Delta\lambda\sim$2000-4000.

The GMOS data reduction pipeline was used to extract and
wavelength-calibrate the spectra of each IFU element.  The variations
in fiber-to-fiber response were removed using twilight flat-fields, and
the wavelength calibration was achieved using a CuAr arc lamp.  We also
applied additional steps to improve the flattening and wavelength
calibration using a series of custom {\sc idl} routines.  The
wavelength coverage of the final data is typically 3900--5500\AA, but
varies depending on exact instrumental setup.  Flux calibration was
carried out using observations of standard stars through the IFU using
identical setups as the target observations.  From the observations of
the standard star, we note that the amplitude of the point-spread
function (PSF) varies by $<$0.05$''$ across the wavelength range
4000-7000\AA\, and so the results in the following sections are
insensitive to variations in PSF with wavelength (0.05$''$ is $\sim$1/4
of the size of a lenslet).

Since the datacubes have a large wavelength coverage we correct for the
parallactic angle in each cube by modeling a 3-hour observation running
from $-1$ to $+1$ hour angle with the corresponding minimum and maximum
airmasses.  After building the datacube, we model and correct this
aberration using a linear interpolation at each slice of the datacube
along the wavelength axis.  We note that the typical airmass of these
observations was $\sim$1.2 and the typical amplitude of the parallactic
angle is $\lsim$0.3$''$ across entire wavelength range of each cube.

\section{Analysis}
\label{sec:analysis}

\subsection{Stellar Masses and Morphologies}

Before discussing the spatially resolved properties, we first make use
of the existing imaging to estimate the stellar masses and morphologies
of the galaxies in our sample.  To estimate the stellar masses, we
follow \citet{McGee11} and use SED modeling employing the Sloan
$ugriz$- and $K$-band photometry \citep{Balogh04}.  Briefly, we
generate galaxy templates using \citet{Bruzual03} models with a
Salpeter initial mass function \citep{Salpeter55} assuming a lower and
upper IMF mass cutoff of 0.1 and 100\,M$_{\odot}$ respectively.  We
follow the galaxy parameter ranges used by \citet{Salim07}.  The model
spectra include a range of galaxy age (0.1--13\,Gyr), metallicity
(0.005--2.5\,Z$_{\odot}$), star formation history and dust obscuration
(A$_v$=0--6\,mags).  In particular, the exponentially declining star
formation rates with superimposed bursts are randomly chosen with a
uniform distribution, and we allow bursts that last some time randomly
distributed in duration between 30 and 300\,Myr.  The strength of the
bursts are also randomly chosen so that during the lifetime of the
burst they produce between 0.03 and 4 times the stellar mass the galaxy
had at the onset of the burst.  From these model star-formation
histories, we generate magnitudes by convolving the resulting spectra
with the $ugriz$ and $K$-band filter response curves.  To fit to the
observations, we search the entire parameter space and minimise the
$\chi^2$, but generate a probability distribution function for each
parameter for each galaxy (see \citealt{McGee11} for a detailed
discussion).  The resulting stellar masses (and their 1$\sigma$ errors)
are given in Table~2.  The median stellar mass of the galaxies in our
sample is 8$\pm$2$\times$10$^{10}$\,M$_{\odot}$.

Since the $K$-band luminosity is a good tracer of the stellar mass
(and is less sensitive to the recent star-formation history than the
optical bands), we can also perform a simple check of the stellar mass
using a canonical mass to light ratio.  \citet{Balogh04} (see also
Fig.~\ref{fig:ugri}) show that the $(u-g)$ and $(r-i)$ colours of E+A
galaxies are consistent with a truncated star-formation model in
which a burst of 5-15\% is superimposed onto a old population, and use
this to derive a canonical M/L$_{K}$=0.8$\pm$0.1.  Applying this to
our $K$-band magnitudes we derive
M$_{*}$=9$\pm$2$\times$10$^{10}$\,M$_{\odot}$, which is consistent
with the estimates from the more sophisticated modeling above.  This
stellar mass is also comparable to the average stellar mass of the
parent sample from \citet{Balogh04};
M$_*=$8$\pm$2$\times$10$^{10}$\,M$_{\odot}$.

The SDSS imaging is also useful for measuring Sersic indices,
asymmetries and radial colour gradients within the galaxies which,
along with the distribution of H$\delta$ equivalent widths, may
provide a diagnostic of the physical mechanism(s) of E+A formation.
For example, positive colour gradients (i.e. bluer in center) can
arise when young stars are more concentrated than the old stellar
population.  In this case, the distribution of H$\delta$ equivalent
widths are also likely to be compact.  In contrast, a negative colour
gradient, together with a positive H$\delta$ equivalent width radial
gradient, may represent a galaxy in which the molecular clouds are not
confined to the nuclear regions.

From our sample of eleven galaxies, three show positive colour
gradients ($\Delta(g-i)/\Delta \log{r}>$0.0), and the rest are either
flat or mildly negative (see also Fig.~\ref{fig:Aextent} and
\citealt{Yang08}).  The galaxies with bluer nuclei (positive color
gradients) may evolve into the negative gradients typical in E/S0s if
the central parts of these galaxies are metal-enhanced \citep{Yang08}.
Finally, we note that the median asymmetry and Sersic index of the
sample is 0.03$\pm$0.01 and 4.9$\pm$1.3 respectively, characteristic of
bulge-dominated systems.

\subsection{Spatial Distributions}
\label{sec:spatial}

To investigate the spatial distribution of the gas and stars in the
galaxies, first we extract narrow-band slices from the datacube around
the emission and absorption lines of interest.  For a given emission
or absorption line, we fit and subtract the continuum using a region
$\pm$100\AA\ from the center of the emission or absorption feature
using a 3--$\sigma$ clip in the fit to be sure that neighboring
emission and absorption lines are omitted.  We also extract continuum
images from the datacube.  For each image, we median filter each
spectral pixel in the datacube between 4350--4750\AA\ (rest-frame) as
a proxy for the ``old'' stellar population (above the 4000\AA\ break)
and between 3650-3850\AA\ (rest-frame) as a proxy for the ``young''
stellar population and show these in the second row of
Fig.~\ref{fig:2Dmaps}.  We caution that the continuum emission from
above and below the 4000\AA\ break has contributions from both ``old''
and ``young'' stars.  For example, for a 10\% by mass starburst which
has been truncated at 1\,Gyr, for solar metallicity the contribution
to the continuum emission from the burst to the total
3650--3850\AA\ and 4350--4750\AA\ continuum emission is $\sim$60\% and
40\% respectively, but can vary from 40\%-70\% in each band depending
on starburst age and metallicity \citep{Maraston98}.

A better test of how the A-stars stars are related to the underlying
stellar emission, (as well as the current star-formation activity) is
to turn to the H$\delta$ absorption line.  We estimate the equivalent
width of H$\delta$ at each spatial pixel following \citet{Balogh99}
(see also \citealt{Goto03}).  We first average each spatial pixel over
a 0.6$\times$0.6$''$ region to approximately match the seeing, and
estimate the continuum flux using $3\sigma$ clipped linear
interpolation between two wavelength windows placed at either side of
the H$\delta$ line (4030\AA\ to 4082\AA\ and 4122\AA\ to 4170\AA;
\citealt{Goto03}).  This 3$\times$3 averaging produces an effective
seeing disk of 0.85$''$, and this is taken into account in all of the
following calculations.  The rest-frame equivalent width of the
H$\delta$ absorption line is then calculated by summing the ratio of
the flux in each pixel of the spectrum, over the estimated continuum
flux in that pixel based on our linear interpolation; its distribution
over the area of each galaxy is shown in Fig.~\ref{fig:centers}.  In
this figure we also indicate the centroids of the nebular emission and
continuum from above the 4000\AA\ break, with a 1$\sigma$ error
ellipse.  To calculate these error ellipses we compute the number of
photons per pixel with their associated $\sqrt{n}$ uncertainties, and
re-compute centroid with 10$^{4}$ monte-carlo realisations.  It is
interesting to note that the regions of strongest H$\delta$ absorption
do not always coincide with regions with the strongest emission lines.
Indeed, Fig.~\ref{fig:centers} shows that, five objects have
H$\delta$, nebular emission and continuum emission which are spatially
co-located on $\sim$\,kpc scales (J0948,J1642,J1627,J1715,J2307),
whilst three systems have significant offsets between nebular emission
and H$\delta$ (J0906,J1242,J1348).  Two galaxies have nebular emission
and H$\delta$ in agreement, but with very extended H$\delta$ which is
offset from the strongest continuum (J0835,J1013).  Thus,
approximately half the sample are (relatively) simple in their
A-star/gas/continuum emission, whilst half are clearly more complex.
On average, the centroid of the A-stars and the continuum emission are
separated by $\Delta r_{\rm H\delta-stars}$=0.8$\pm$0.4\,kpc, whilst
the average offset between the H$\delta$ and nebular emission is
$\Delta r_{\rm H\delta-gas}$=1.0$\pm$0.2\,kpc, but can be as large as
as 1.7\,kpc.  This suggests that at least in some cases, the new OB
stars are either displaced from the older A-stars, or that the
extinction is not uniform.  In either case this helps to reconcile the
fact that both emission and strong absorption can be seen in the same,
integrated galaxy spectrum.  The two populations do not generally
arise from exactly the same place.

Next, we examine the spatial extent of the A-stars through the
H$\delta$ absorption.  In the following calculations, we assume that
both the continuum emission is smoothly distributed and subtract the
amplitude of the point-spread function at the redshift of the galaxy
in quadrature from the area's covered by the continuum and A-stars.
As shown by the distributions in Fig.~\ref{fig:centers}, the A-stars
are typically distributed over a large radius.  We quantify this by
computing the radius at which the H$\delta$ equivalent width drops to
half the peak value (this is extracted along the major kinematic axes
within the galaxy which is discussed in \S~\ref{sec:kinematics}).
This quantity, $r_h(W_\circ H\delta)$ is compared with the observed
$g-i$ colour gradient within the central 2\,kpc in the top-left panel
of Fig.~\ref{fig:Aextent}.  Positive radial colour gradients indicate
galaxies that are bluer in the center, and are measured in only three
galaxies ($\sim$30\%).  This is lower than the fraction of E+A
galaxies with strong positive colour gradients (i.e. blue cores) found
by \citet{Yamauchi06} ($\sim$67\%), but still higher than the fraction
of early type galaxies with positive colour gradients
\citep{Balcells94}.  Within our sample, there is no evidence for the
correlation between colour gradient and spatial extent of the A-stars
(as might have been expected from merger or tidal models for E+A
formation; \citealt{Bekki05}).

Another way to characterise the extent of the H$\delta$ absorption
feature is to measure the area over which the equivalent width exceeds
some threshold.  We choose a threshold of 6\AA, and calculate the
physical area A(H$\delta$\,$>$\,6\AA) for each galaxy.  The results
are shown in Fig.~\ref{fig:Aextent} as a function of various kinematic
parameters, discussed below.  Note that in all but one galaxy
(J0948+0230) the area with \WoHd$>6$\AA\ is significantly larger than
the seeing.  The median A(H$\delta$\,$>$\,6\AA)=4.0$\pm$2.1\,kpc$^2$,
with a range of A(H$\delta$\,$>$\,6\AA)=0.8--15.2\,kpc$^2$.  In
comparison to the continuum emission from above the 4000\AA\, break,
the A-stars cover at least $\sim$10\% of the area, with a median of
33$\pm$5\%.  Such wide-spread A-star populations have also been found
in previous samples: \citet{Goto08a} show that the A-stars in two SDSS
selected E+A galaxies are distributed over $\sim$10\,kpc$^2$, while
the seven E+A galaxies from \citet{Pracy09} have A-stars that are
typically distributed over 5--15\,kpc$^2$.

In summary, we note two important observations.  First, the H$\delta$
is widespread, typically over
A(H$\delta$\,$>$\,6\AA)=4.0$\pm$2.1\,kpc$^2$ and thus the unusual
spectrum characteristic of E+A galaxies is a property of the galaxy as
a whole.  It is not due to a heterogeneous mixture of populations,
such as a nuclear starburst with a very high equivalent width
($>$10\AA) on top of an otherwise normal galaxy.  Second, in
approximately half of our sample, the A-stars, nebular emission and
continuum are not co-located, suggesting that the newest stars are
often forming in a different place than those that formed
$\lsim$1\,Gyr ago, and that recent star-formation has occurred in
regions distinct from the continuum emission (the older stellar
populations).

\subsection{Kinematics}
\label{sec:kinematics}

To obtain the stellar velocity structure within the galaxies we first
construct a continuum image of the galaxy from the datacube collapsed
between 4350--4750\AA, and bin the data to a constant signal-to-noise
ratio per bin using the Voronoi binning method \citep{Cappellari03}.
We demand a signal-to-noise of 30 in continuum, and then cross
correlate each of the individual spatial regions with a library of
template spectra \citep{Vazdekis99}.  The cross correlation is
performed using the penalized pixel fitting method ({\sc ppxf};
\citealt{Cappellari04}).  To maximise signal-to-noise, to derive the
velocity field of the young stellar population we use the wavelength
range $\sim$3800-4290, which covers Ca H\&K and H$\delta$.

As Fig.~\ref{fig:2Dmaps} shows, all eleven galaxies have velocity
gradients in their dynamics, with amplitudes ranging from
30--200\,km\,s$^{-1}$.  At least eight have dynamics that appear to be
consistent with rotation; i.e., the peak of the line of sight velocity
dispersion is at the dynamical center, and the velocity field is
characteristically bisymmetric, showing a so-called ``spider'' diagram.
For these galaxies, we model the velocity field in order to estimate
the inclination and true rotational velocity.  To achieve this, we use
a simple approach and model the velocity field with an {\it arctan}
function to describe the shape of the rotation curve such that $v(r) =
v_c\,{\mbox arctan}(r/r_t)$, where $v_c$ is the asymptotic rotational
velocity and $r_t$ is the effective radius at which the rotation curve
turns over \citep{Courteau97}.  We then construct a two-dimensional
kinematic model with six free parameters ($v_c$, $r_t$, [x/y] center,
position angle (PA) and inclination) and use a genetic algorithm with
10$^{5}$ random initial values and a scale factor of 0.95 to search for
a best fit.  We demand $>$30 generations are performed before testing
for convergence to a solution, throwing out (a maximum of) 10\% of the
solutions in each generation.  The best fit kinematic maps are shown as
contours in Fig.~\ref{fig:2Dmaps}, and the best-fit inclination and
rotation speed are given in Table~2.  Using the velocity maps and model
fits, we identify the major kinematic axis wherever possible and
extract the one dimensional rotation curve and velocity dispersion
profile along and show these in Fig.~\ref{fig:1dprofiles}.  From each
of the two-dimensional maps, we also extract the distribution of
H$\delta$ equivalent widths, both in radial bins and along the
kinematic cross section.

\begin{figure}
\centerline{
  \psfig{figure=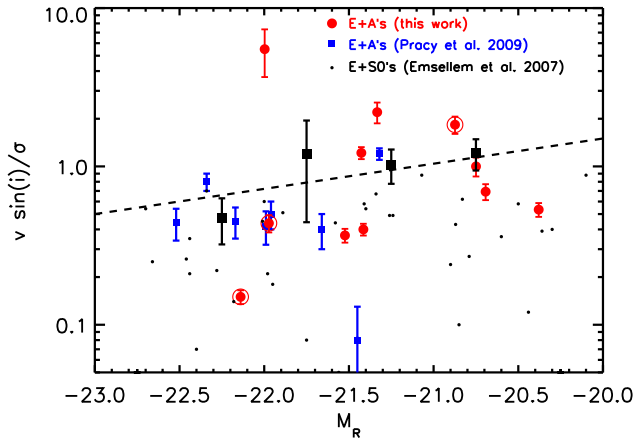,width=3.5in,angle=90}}
\caption{The dynamical measurement of $v\sin(i)/\sigma$ as a function
  of absolute $R$-band magnitude for E+A galaxies from our sample and
  the recent work by \citet{Pracy09}.  The solid black points denote
  the average values in bins of $\Delta m$=0.5 and the dashed line
  shows a linear fit, indicating a weak trend such that fainter
  galaxies tend to have higher $v\,sin(i)/\sigma$.  The plot also
  includes a comparison to the early-type and elliptical galaxies from
  \citet{Emsellem07} (we have colour corrected their sample from
  M$_{B}$ to M$_{R}$ using a $(B-R)$=2.0 appropriate for an early type
  galaxy at $z$=0.05).  This sample have a lower ($v/\sigma$), with
  $v\,sin(i)/\sigma$ (0.30$\pm$0.03), but show a comparably weak trend
  of with increasing absolute magnitude.  Galaxies which are weak
  radio sources (and thus may harbor AGN are shown as open symbols);
  do not stand out from the rest of the sample.}
\label{fig:Mrvs}
\end{figure}

\begin{figure*}
\centerline{
  \psfig{figure=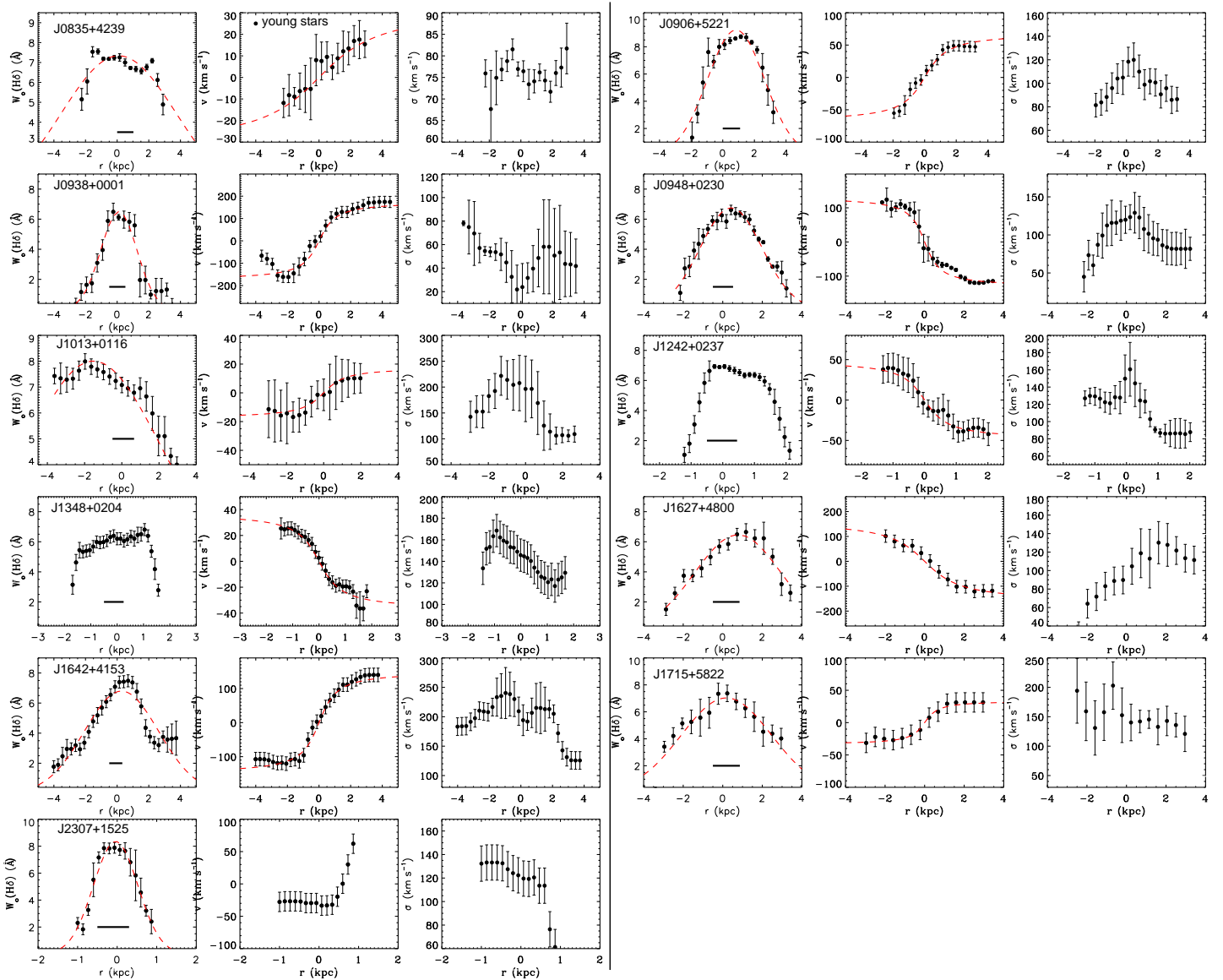,width=7.5in,angle=0}}
\caption{Extracted, one dimensional equivalent width, velocity and line
  of sight velocity dispersion profiles are shown for each of the E+A
  galaxies in our sample.  The profiles are extracted along the
  major axis as indicated in Fig.~\ref{fig:2Dmaps}.  {\it Left:} One
  dimensional extent of equivalent widths. {\it Middle:} One
  dimensional rotation curve of the galaxy with the best fit rotational
  model (where appropriate) overlaid as contours {\it Right:} One
  dimensional distribution of line widths.  The solid bar in the left
  hand panel denotes the amplitude of the seeing disk, converted to
  physical scale at the redshift of each galaxy.}
\label{fig:1dprofiles}
\end{figure*}

\begin{figure*}
\centerline{
  \psfig{figure=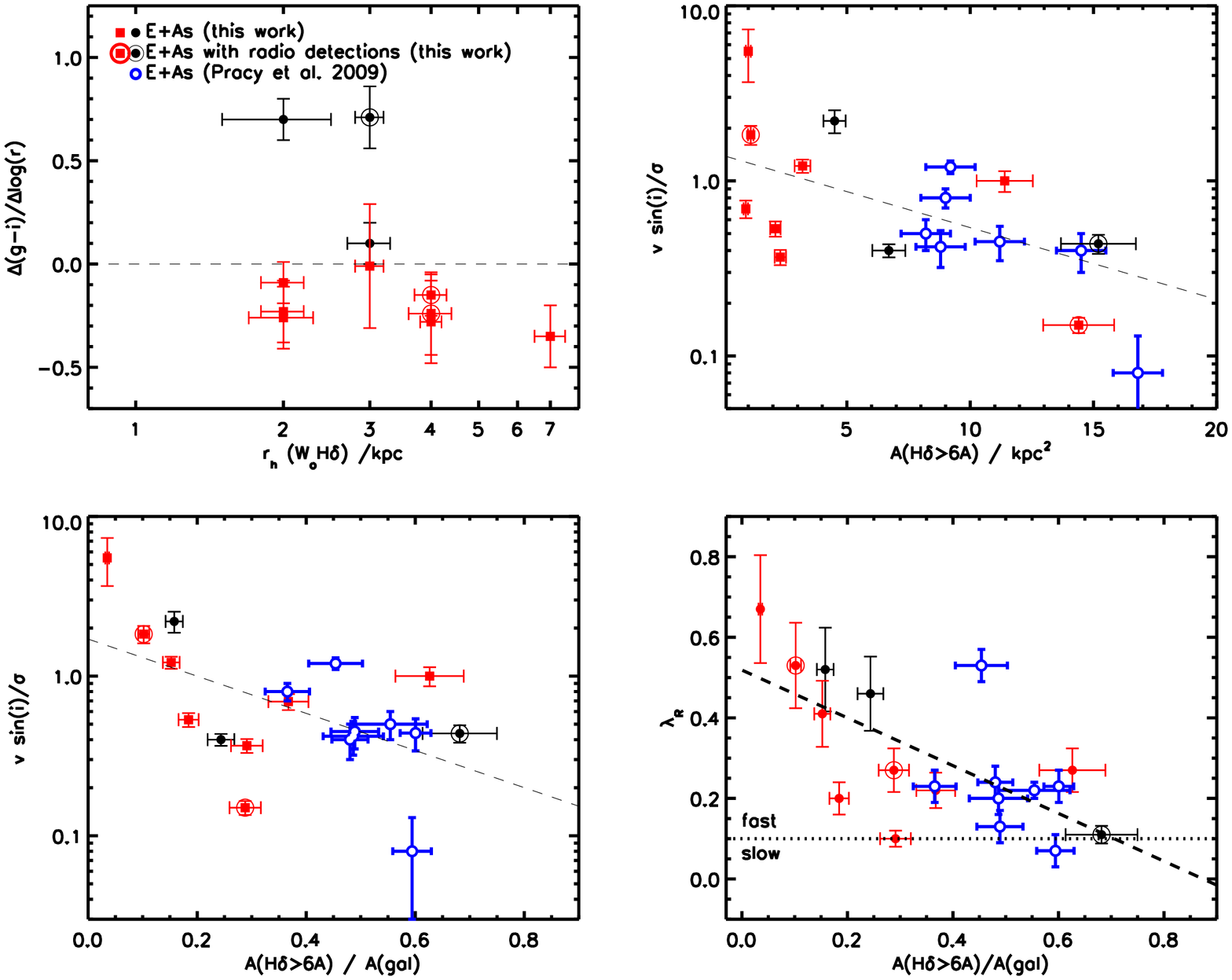,width=6.0in,angle=90}}
\caption{The spatial extent of the A-stars within the E+A galaxies in
  our sample as a function of colour gradient and kinematics.  {\it
    Top Left:} The extent of the A-stars (defined as the radius at
  which the H$\delta$ equivalent width drops to half the peak value
  along the major kinematic axis) as a function of the $g-i$ colour
  gradient within the central 2\,kpc of the galaxy.  Positive radial
  colour gradients denote light that is bluer in the center.  The
  dashed line denotes a colour gradient of zero, and we use this to
  differentiate galaxies with positive- (black filled circles) from
  negative- colour gradients (red filled squares).  In all panels we
  identify the E+As which have weak radio emission
  (L$_{1.4}\gsim$10$^{22}$\,W/Hz; highlighted with an open circle).
  {\it Top Right:} The spatial extent of the A-stars as a function of
  the dynamics as measured from ($v\,sin(i)/\sigma$) for the E+A
  galaxies in our sample.  We also include the six E+As from
  \citet{Pracy09}.  To provide a quantitative measure of the spatial
  extent of the A-stars, we define A(H$\delta>6$\AA) as the area over
  which the H$\delta$ is stronger than 6\AA.  The galaxies with higher
  $v\,sin(i)/\sigma$ tentatively show signs of having more compact
  A-star distributions compared to galaxies which are pressure
  supported; the average $v\sin(i)/\sigma$ for the galaxies where
  A(H$\delta$$>$6$\AA$)$>$4\,kpc$^{2}$ is 0.5$\pm$0.2, compared with
  $v\,sin(i)/\sigma$=1.1$\pm$0.5 for the others.  {\it Bottom Left:}
  Since the measurement of the spatial extent of H$\delta$ is
  sensitive to detecting the continuum, we normalise the area,
  A(H$\delta$), by the spatial extent of the galaxy, as measured from
  the continuum above the 4000\AA\ break.  We note that the results
  are insensitive to whether we use the SDSS r-band image to measure
  the half light radius, or R$_{e}$.  This plot shows that the
  galaxies with the most compact H$\delta$ distribution tend to be
  those which have dynamics consistent with rotation, while the
  galaxies with A-stars which are most wide-spread are those with
  dispersion dominated dynamics. {\it Bottom Right:} The spatial
  extent of the A-star population normalised to the continuum as a
  function of $\lambda_R$ \citep{Emsellem07}.  High and low
  $\lambda_R$ correspond to `fast' and `slow' rotators respectively.
  This plot shows that the fastest rotators tend to have the most
  compact A-star spatial distribution.  We note that although our
  sample is small, the galaxies with weak AGN do not stand out in
  terms of their A-star extent or dynamics from the rest of the sample
  in any of the panels.}
\label{fig:Aextent}
\end{figure*}

A measure of whether the dynamics are dominated by ordered- or random-
motion can be obtained by measuring the ratio of rotational-velocity to
line of sight velocity-dispersion, $v\,\sin(i)/\sigma$.  We derive a
median luminosity-weighted $v\,\sin(i)/\sigma$=0.8$\pm$0.4 for the
whole sample with a range of 0.2--5.5.  However, Fig.~\ref{fig:Mrvs}
shows that there is a weak correlation between absolute magnitude and
$v/\sigma$, such that the lower luminosity systems have an increasing
rotational support, although only four of the eleven galaxies have
$v/\sigma>1$, uncorrected for inclination.  Recent studies of early
type galaxies from \citet{Emsellem07} identify a comparably weak trend
of v/$\sigma$ with magnitude (although we note that their sample has a
slightly lower average v$/\sigma$ with an average
v/$\sigma$=0.30$\pm$0.03).  Indeed, the early types in
\citet{Emsellem07} have M$_{R}$=-21.8$\pm$0.3 compared to
M$_{R}$=-21.5$\pm$0.5 for the E+As, and with stellar masses of
1.0$\pm$0.3$\times$10$^{11}$\,M$_{\odot}$, which is also consistent
with our sample.  Thus, these early-types and ellipticals from
\citet{Emsellem07} also have many of the properties expected for the
descendant population of E+A galaxies.

\citet{Bender94} also find a correlation between magnitude and
v/$\sigma$ for luminous elliptical galaxies.  Although the
\citet{Bender94} sample are dominated by brighter galaxies than the
E+A's studied here (typically $\sim$1.3\,mags brighter), those
galaxies with high v/$\sigma$ ($\sim$1 and so comparable to the E+A's)
are increasing dominated by lower mass, lower luminosity Sb--S0s.

A better test of the dynamical state of galaxies with low v/$\sigma$
can be made by measuring the luminosity-weighted stellar angular
momentum per unit mass, $\lambda_R$ as developed by
\citet{Emsellem07}.  This combines the velocity amplitude and line of
sight velocity dispersion
via:\\ $\lambda_R=<R|v|>/<R\sqrt{v^2+\sigma^2}>$, where $R, v$ and
$\sigma$ are the radius, rotational velocity and velocity dispersion
from the kinematic center.  $\lambda_R$ is designed to quantitatively
distinguish galaxies that have similar $v/\sigma$ values but very
different velocity structures by identifying those galaxies with
significant angular momentum per unit mass.  \citet{Emsellem07} define
``fast'' and ``slow'' rotators as $\lambda_R>$0.10 and
$\lambda_R<0.10$ respectively.  Classically, ``fast'' rotators are
dominated by low mass spheroids (with a large v/$\sigma$) whilst
``slow'' rotators are dominated by ellipticals, albeit with a large
range in M$_R$.  We report $\lambda_R$ for each galaxy in Table~3.
Our sample has a median $\lambda_R$=0.35$\pm$0.1, and all of the
galaxies in our sample have $\lambda_R>0.1$.  When including the E+A's
from \citet{Pracy09}, 18/19 galaxies with well resolved dynamical maps
have $\lambda_R>0.1$.  Taken with the measurements of v/$\sigma$,
$\lambda_R$ and stellar masses, the E+As in our sample have properties
comparable to those measured in local, representative elliptical
galaxies and S0s where similar measurements have been made.

To investigate how the dynamics and A-star populations are related, we
correlate the kinematics as measured from $v/\sigma$ and $\lambda_R$
as a function of the A-star spatial extent and colour gradients.
Fig.~\ref{fig:Aextent} shows that the galaxies which have the highest
v/$\sigma$ also have the most compact A-star distributions.  Indeed,
the average $v/\sigma$ for the galaxies where
A(H$\delta>6$\AA)$>$4\,kpc$^{2}$ is 0.5$\pm$0.2, compared with
$v/\sigma$=1.1$\pm$0.5 for the remainder.  We note that we would
expect the opposite correlation between the spatial extent of the
A-stars and v/$\sigma$ if the measurements were being driven by
signal-to-noise effects alone (low v/$\sigma$ values would be expected
for the most compact sources where it is more difficult to identify
large velocity gradients).  However, measuring the spatial extent of
H$\delta$ absorption is sensitive to detecting the continuum, so we
also normalise the area with A(H$\delta>$6\AA) by the spatial extent
of the continuum emission, A(gal), as measured from the continuum.
For the continuum measurement we use the flux redward of the
$4000$\AA\ break as measured from our IFU observations, although we
note that the results are quantitatively similar when using the r-band
half light radius, or R$_{e}$, measured from SDSS imaging.  In all
panels, we find that galaxies with high v/$\sigma$ show the most
compact A-star populations.

Of course, the spatial distribution of an equivalent width depends on
its intrinsic spatial distribution as well as the overall distribution
of the continuum light (or galaxy surface brightness profile) after
convolving with the seeing \citep[e.g.\ ][]{Pracy05}.  In particular,
for a galaxy with a centrally concentrated equivalent width profile,
the steeper the continuum surface brightness profile, the more the
stellar population in the galaxy center will contaminate the outer
regions after convolution with the seeing (see \citealt{Pracy10} for a
detailed discussion).  In principle this could give a correlation
between v/$\sigma$ and the extent of the A-star light distribution.
Whilst full modelling to recover the intrinsic equivalent width
distribution is beyond the scope of this paper, we can perform a
simple test of whether this effect is likely to significantly
contribute our results.  We therefore construct 10$^4$ mock GMOS
datacubes with variable amplitude velocity fields, continuum emission-
and H$\delta$ equivalent width- profiles (which broadly bracket the
range of our observations) and convolve these with the seeing.  We
note that in these simulations, the extent of the continuum emission
and H$\delta$ profiles are independent.  We measure the ``true''
v/$\sigma$ and ratio of A(H$\delta>$6)/A(gal) as well as the same
quantities from the recovered cubes after convolution with the seeing
and surface brightness limits appropriate for our observations.
Allowing the spatial extent of the A-stars to vary from ``unresolved''
to 3$\times$ the continuum emission, we find that the change in slope
of $\Delta log(v/\sigma)$/$\Delta(A(H\delta>$6)/A(galaxy)) is -0.04 to
-0.2 depending on choice of parameters.  The gradient of the
correlation seen in Fig.~\ref{fig:Aextent} is $\Delta
log(v/\sigma)$/$\Delta(A(H\delta>$6)/A(galaxy))=-2.2$\pm$0.3.  Thus
the underlying A-star distribution, continuum surface brightness
profile and seeing may contaminate the correlation between the extent
of the A-stars and dynamics seen in Fig.~\ref{fig:Aextent}, but it
seems unlikely that this effect alone can account for the correlation
we see.  We also note that in general, rotation-dominated galaxies
have shallower surface brightness profiles than pressure-supported
galaxies \citep[e.g.\ ][]{Courteau97}.  This means there is a physical
correlation between the two which may lead to an indirect correlation
between v/$\sigma$ and A(H$\delta$)/A(gal).  However, clearly,
observations of a well matched control sample of non E+A's around the
H$\delta$ are required to examine whether the correlation in
Fig.~\ref{fig:Aextent} is a result of the processes which drive the
E+A's.  Nevertheless, for our sample the `slow-rotators' tend to have
the most wide-spread A-star populations, while the `fast-rotators'
have the more compact A-star distributions.

\begin{table*}
{\tiny
\begin{center}
{\centerline{\sc Table 2: Properties of the E+A Galaxies in Our Sample}}
\smallskip
\begin{tabular}{lccccccccccccc}
\hline
\noalign{\smallskip}
                  &   \multicolumn{2}{|c|}{Equivalent Width}& B/T & Sersic  & M$_{*}$              & [O{\sc ii}] Flux      & SFR([O{\sc ii}])         & $v$\,$sin(i)/\sigma$ & inc $(i)$  & $v_c$      & $\lambda_R$   & $r(H\delta)$   & $A_{H\delta}>6$\AA\  \\
                  &   [O{\sc ii}] & H$\delta$     &           & index       & ($\times10^{10}$     & ($\times$10$^{-16}$   & (M$_{\odot}$/yr            &                    & (deg)        & (km/s)     &               & (kpc)          & (kpc$^2$)               \\
                  & (\AA)         & (\AA)         &           & $(n)$       & M$_{\odot}$)         & erg/s/cm$^2$)        &                           &                    &              &            &               &                &                   \\   
\hline                                                                                                                                                                                                                                                       
J0835+4239       & 2.9$\pm$0.2  & 6.9$\pm$0.3    & 0.48      & 7$\pm$1     & 8.3$_{-1.9}^{+2.6}$   & 5.0$\pm$0.5           & 0.14$\pm$0.07             & 0.4$\pm$0.1        & 51$\pm$11    & 55$\pm$15  & 0.11$\pm$0.04 & 2.4$\pm$0.2    & 15.2               \\
J0906+5221$^{*}$ & 1.5$\pm$0.3  & 6.3$\pm$0.4    & 0.69      & 4$\pm$1     & 2.0$_{-0.3}^{+0.5}$   & 2.3$\pm$1.0           & 0.08$\pm$0.04             & 1.0$\pm$0.1        & 26$\pm$5     & 75$\pm$12  & 0.27$\pm$0.04 & 2.0$\pm$0.2    & 11.4               \\
J0938+0001       & 3.9$\pm$0.4  & 5.0$\pm$0.3    & 0.47      & 3.5$\pm$1.0 & 23.4$\pm$6.0        & 5.1$\pm$0.5           & 0.15$\pm$0.04             & 5.5$\pm$0.1        & 57$\pm$6     & 202$\pm$12 & 0.67$\pm$0.05 & 1.2$\pm$0.5    & $>$0.8               \\
J0948+0230       & 2.4$\pm$0.5  & 6.4$\pm$0.2    & 0.29      & 2.2$\pm$0.3 & 6.6$_{-1.2}^{+1.7}$   & 3.8$\pm$0.9           & 0.05$\pm$0.03             & 1.8$\pm$0.1        & 54$\pm$7     & 200$\pm$10 & 0.53$\pm$0.03 & 1.2$\pm$0.4    & $>$0.8               \\
J1013+0116       & 5.2$\pm$0.3  & 6.8$\pm$0.5    & 0.46      & $>$7        & 15.0$_{-2.8}^{+7.2}$   & 12.1$\pm$0.2          & 0.48$\pm$0.12             & 0.2$\pm$0.1        & 42$\pm$10    & 152$\pm$14 & 0.27$\pm$0.04 & 2.8$\pm$0.3    & 14.4               \\
J1242+0237$^{*}$ & 2.0$\pm$0.4  & 6.0$\pm$0.2    & 0.40      & 3.8$\pm$0.5 & 1.9$_{-0.2}^{+0.5}$   & 2.7$\pm$0.6           & 0.06$\pm$0.03             & 0.5$\pm$0.1        & 21$\pm$5     & 92$\pm$6   & 0.20$\pm$0.03 & 1.0$\pm$0.2    & 2.1                \\
J1348+0204       & 2.1$\pm$0.5  & 6.0$\pm$0.3    & 0.67      & 6.2$\pm$0.4 & 6.9$_{-1.3}^{+1.6}$   & 7.1$\pm$0.8           & 0.11$\pm$0.03             & 0.4$\pm$0.1        & 49$\pm$7     & 63$\pm$8   & 0.10$\pm$0.04 & 1.6$\pm$0.2    & 2.2                \\
J1627+4800       & 0.8$\pm$0.4  & 5.2$\pm$0.5    & 0.73      & $>$7        & 5.7$_{-1.2}^{+1.5}$   & $<$2.1                & $<$0.13                   & 2.2$\pm$0.1        & 37$\pm$5     & 229$\pm$12 & 0.52$\pm$0.05 & 2.1$\pm$0.5    & 4.4                \\
J1642+4153       & 1.2$\pm$0.5  & 6.5$\pm$0.5    & 0.64      & 3.5$\pm$0.6 & 6.0$_{-1.1}^{+1.7}$   & 4.5$\pm$1.0           & 0.08$\pm$0.03             & 1.2$\pm$0.1        & 51$\pm$6     & 201$\pm$10 & 0.41$\pm$0.04 & 1.5$\pm$0.3    & 3.2                \\
J1715+5822       & 2.2$\pm$0.4  & 6.5$\pm$0.8    & 0.48      & 4.5$\pm$1.0 & 7.9$_{-1.9}^{+2.5}$   & 2.8$\pm$0.5           & 0.17$\pm$0.05             & 0.4$\pm$0.1        & 49$\pm$5     & 90$\pm$5   & 0.46$\pm$0.03 & 2.7$\pm$0.3    & 6.5                \\
J2307+1525       & 1.7$\pm$0.3  & 6.3$\pm$0.4    & 0.45      & $>$7        & 4.5$_{-0.9}^{+1.3}$   & 3.0$\pm$0.5           & 0.05$\pm$0.02             & 0.7$\pm$0.1        & ...          & ...        & 0.22$\pm$0.03 & 1.4$\pm$0.2    & 2.8                \\
\hline\hline
\label{table:props}
\end{tabular}
\vspace{-0.5cm}
\caption{Notes: The two E+A galaxies labeled with a $^{*}$
  (J0906+5221 and J1242+0237) are both in cluster environments.  $n$
  denotes the Sersic index measured from the SDSS $gri$ photometry.
  The estimate of the star-formation rate (SFR) is calculated using the
  [O{\sc ii}]$\lambda3727$ emission line flux and assuming the
  calibration from \citet{Kennicutt98}.}
\end{center}
}
\end{table*}

\section{Discussion}

There are at least three important unresolved problems regarding E+A
galaxies: 1) from where in the galaxy does the unusual spectrum
originate? 2) what star formation history leads to the spectral
characteristics of these galaxies; 3) what triggers the recent change
in star formation history?  To address these issues, we have performed
three dimensional spectroscopy of a sample of eleven massive E+A
galaxies selected from the SDSS for their unusually strong H$\delta$
equivalent widths but weak [O{\sc ii}] emission, suggesting that star
formation in these galaxies was recently truncated, possibly following
a starburst.

Our sample was selected to span a range of morphology and environment
and hence is diverse by design.  Moreover, with a limited sample size
of evelen galaxies it is not possible to answer any of these questions
definitively, especially since there may be more than one way to form
an E+A galaxy.  Nevertheless, there are several important results that
are generic within our sample.  First, we note that we see no strong
correlation of the galaxy dynamics with the $K-$band
morphology, or the environment of the galaxy with their kinematics or
morphology of the nebular emission or A-stars: the two E+A galaxies in
dense environments do not appear to stand out from the rest of the
sample in any of their properties, so we discuss the properties of the
ensemble below.

To investigate the likely age of the starbursts and hence
star-formation history, we estimate the age of the A-star populations
by comparing the average galaxy spectrum for each galaxy with the
spectral library of \citet{Jacoby84}.  We use the strength of the Ca
K\,3933\AA\ absorption line (which is especially sensitive to age;
\citealt{Rose85}), together with the equivalent width of the Ca
H\,3969\AA\ (and H$\epsilon$) absorption lines.  For all of the E+A
galaxies in our sample, the spectra best resemble that of an A5$\pm$1
star, with an effective temperature of 8160\,K.  Interpolating the
theoretical isochrones from \citet{Bertelli94}, this suggests a
luminosity weighted age of 0.8$\pm$0.1\,Gyr (for the MS turn off at A8)
although we caution that stellar mixes and metallicity make this
estimate uncertain.  Nevertheless, this is also supported by the
broad-band colours which are well matched by a model in which an
instantaneous starburst (which accounts for 10\% by mass), is
superposed upon the continuum above the 4000\AA\ break
(Fig.~\ref{fig:ugri}).

Thus, the colours and line strengths we observe can best be reproduced
with $\sim 10$ per cent (by mass) starbursts on top of an old
population \citep[e.g.][]{Balogh04,Shioya04}.  It is also evident that
any residual star formation in the galaxy is negligible: the median
stellar mass of the E+A galaxies in our sample is
8$\pm$2$\times$10$^{10}$\,M$_{\odot}$, suggesting an average burst
mass of M$_{\rm burst}\sim$1.0$\times$10$^{10}$\,M$_{\odot}$.  The
median star-formation rate of galaxies in our sample (including
limits) is SFR([O{\sc ii]})=0.13$\pm$0.03\,M$_{\odot}$\,yr$^{-1}$,
less than 2\% of the past average star-formation rate.  Thus, the
presence of even moderately strong H$\delta$ absorption ($>3$\AA\ or
so) indicates that the star-formation rate at truncation was at least
comparable to the past average rate.  Integrating the current
star-formation rate over 1\,Gyr we derive a mass of
$\sim$1$\times$10$^{8}$\,M$_{\odot}$, or only $\sim 0.1$ per cent of
the total galaxy mass.  Since the mass involved in the burst is at
least $\sim$10\% of the mass of the stellar mass, thus suggests that
the main progenitors of our E+A galaxies had significantly higher
star-formation rates than seen today, and were most likely blue-cloud,
star-forming systems.

The A-stars are widely distributed, with an average
A(H$\delta$\,$>$\,6\AA )=4.0$\pm$2.0\,kpc$^2$, and a range of
A(H$\delta$\,$>$6\AA )=0.8--15.2\,kpc$^2$ (see also
\citealt{Swinbank05a,Goto08a,Pracy09}).  On average, the strongest
H$\delta$ (W$_{o}$(H$\delta$)\,$>$6\AA\,) covers 33$\pm$5\% of the area
defined by the detected continuum above the 4000\AA\ break.  It is also
interesting that none of our galaxies show regions with very strong
absorption, W$_{o}$(H$\delta$)$>$10\AA, at least on $\sim$kpc scales.
Such strong absorption would be an unambiguous indication of a
significant {\it enhancement} of the star formation rate in the past
$\lsim 1$ Gyr; weaker lines can be formed via truncation of star
formation in normal, gas-rich galaxies \citep[e.g.][]{Balogh99}.
Moreover, it is also clear that in approximately half of the sample,
the A-stars (which give rise to the absorption) and nebular emission
are not co-located, suggesting that the newest stars are often forming
in a different place than those that formed $\lsim1$Gyr ago.  Thus,
together with the burst mass to stellar mass ratio, this suggests
that there needs to have been a recent, dramatic change in star
formation history, coordinated over kiloparsec scales in these
galaxies.

The galaxies display a range of kinematic properties: all eleven have
velocity gradients in their stellar kinematics and we derive a median
luminosity-weighted $v\,\sin(i)/\sigma$=0.8$\pm$0.4 for the whole
sample (with a range of 0.2--5.5).  Fig.~\ref{fig:Mrvs} shows that
there is a weak correlation between absolute magnitude and $v/\sigma$,
such that the lower luminosity systems have an increasing rotational
support.  The median $\lambda_R$ for the E+A's in our sample is
$\lambda_R$=0.35$\pm$0.1, and all of the E+A galaxies studied here have
$\lambda_R>$0.1 which is consistent with that found for a comparably
bright sample of elliptical galaxies and passive S0's which are also
well matched in their average stellar masses \citep{Emsellem07} .

Although we have shown that the E+A phase is a galaxy-wide phenomenon,
conclusively addressing what triggers the E+A phase is more difficult,
especially since there are a number of processes which could be
responsible at once (e.g.\ strong interactions or mergers, possibly
proceeded by AGN activity).  However, we can search for global trends
in the A-star population with other galaxy properties in order to
search for the major contributing factor.  We find that dynamics are
correlated with the spatial extent of the A-stars such that the
`slow-rotators' have the most wide-spread A-star populations, while the
`fast-rotators' have compact A-star distributions.  Since the E+A
galaxies in our sample do not preferentially lie in dense/cluster
environments, it is unlikely that any truncatation of star-formation
was due to ram pressure stripping of the gas.

It is interesting to note that theoretical models which aim to trace
the formation and evolution of early type galaxies have shown that
there are two primary factors which determine the the structure of
early-type galaxies.  More anisotropic and slowly rotating galaxies
result from predominantly collisionless major mergers, while faster
rotating galaxies are produced by more gas-rich mergers, where
dissipation plays an important role
\citep{Kormendy89,Bender92,Faber97}.  The second major factor which
drives the galaxy structure is the mass fraction of the merger
components.  Unequal-mass galaxy mergers tend to produce more rapidly
rotating galaxies than mergers which comparable mass ratios
\citep{Naab99,Naab03}.  Whichever process was responsible, the
encounter has left residual rotational motion in at least 90\% of our
sample, with A-stars widely distributed within the ISM.
Qualitatively, our results are consistent with models of unequal mass,
gas rich mergers.

Finally, it is interesting to examine whether
activation of a central black hole could be related to the end of star
formation \citep{Yan06}.  As described in \S~\ref{sec:gmos}, four of
our galaxies are detected in FIRST with 1.4\,GHz fluxes in the range
0.5--2.2\,mJy, corresponding to luminosities of
L$_{1.4}$=8--15$\times$10$^{21}$\,W\,Hz$^{-1}$.  All four galaxies with
FIRST detections show [O{\sc iii}] and (weaker) [O{\sc ii}] emission.
For these galaxies, the centroid of the nebular emission appears offset
from the continuum by 0.4--1\,kpc, but this is consistent with the rest
of the sample.  Thus, if there is an AGN in these galaxies, it does not
appear to be the sole contributor to the nebular emission.  Moreover,
if the radio emission were due to star formation, then the
instantaneous star-formation rate would be 6--11M$_{\odot}$\,yr$^{-1}$
\citep{Helou85} a factor $\sim$60 times that inferred from the nebular
emission lines.  This could be possible if the galaxies are
exceptionally dusty, although this seems unlikely given the blue
$(r-i)$ colours; we would expect dusty, star-forming galaxies to lie on
the upper edge of the blue cloud in Figure~\ref{fig:ugri}, with
$(r-i)>0.4$ \citep{Wolf05,Balogh09}.  Thus it is most likely that the
radio emission arises from an AGN.  In all the key figures we have
highlighted the E+A galaxies in our sample that have
L$_{1.4}\gsim$10$^{22}$\,W/Hz.  In all cases their properties appear to
be consistent with those of the rest of the sample.  Thus, although we
only have small number statistics, our results suggest that AGN
activity in E+A galaxies does not play a dramatic role in defining the
spatial or kinematic properties of the A-stars, and/or that its effects
are short--lived.

\section{Conclusions}

We have investigated the spectro-photometric properties of a sample of
eleven E+A galaxies in the redshift range $0.12<z<0.07$ in order to
provide constraints on the physical processes which cause the unusual
post-starburst signatures.  Although our sample selection was
hetrogeneous (by design we selected E+A's with a range of morphologies
and environments), there are several important results which can be
summarized as:

$\bullet$ The strongest H$\delta$ ($W_o>$6\AA) tends to widely
distributed within the galaxies, on average covering $\sim33$\% of the
galaxy image and extending over areas of 1--15\,kpc$^2$.  This
suggests that the characteristic E+A signature is a property of the
galaxy as a whole and not due to a heterogeneous mixture of stellar
populations.

$\bullet$ In approximately half of the sample, the A-stars, nebular
emission and continuum are not co-located.  The average offset between
the H$\delta$ and nebular emission is $\Delta r_{\rm
  H\delta-gas}$=1.0$\pm$0.2\,kpc, but can be as large as as 1.7\,kpc.
This suggests that at least in some cases, the new OB stars are either
displaced from the older A-stars, or that the extinction is not
uniform.  In either case this helps to reconcile the fact that both
emission and strong absorption can be seen in the same, integrated
galaxy spectrum.  The two populations do not generally arise from
exactly the same place.

$\bullet$ The colours and line strengths of these E+A galaxies are
consistent with a $\sim 10$ per cent (by mass) starburst on top of an
old population suggesting an average burst mass of M$_{\rm
  burst}\sim$10$^{10}$\,M$_{\odot}$.  The current (residual)
star-formation is unable to account for the burst mass.  This
indicates that the main progenitors of most of our E+A galaxies were
blue-cloud, star-forming galaxies in which there was a recent,
dramatic change in star formation history, coordinated over kiloparsec
scales.

$\bullet$ The kinematics show that some level of rotation in the
A-star population is common, and the dynamics (v/$\sigma$ and
$\lambda_R$) are consistent with those measured in comparably bright
early types recently studied by \citet{Emsellem07}, which are also
well matched in their average stellar masses.   Whichever
process was responsible for causing the starburst and subsequent
truncation, the encounter has left residual rotational motion in at
least 90\% of the sample, with A-stars widely distributed within the
ISM.

$\bullet$ We find that the A-star dynamics are correlated with their
spatial extent such that the `slow-rotators' have the most wide-spread
A-star populations, while the `fast-rotators' have compact A-star
distributions.

$\bullet$ We also find that the fraction of galaxies which have radio
emission suggestive of low luminosity AGN is 20--40\%, a factor
$\sim$8$\times$ higher than expected given their stellar masses,
indicating a high AGN fraction of E+A galaxies.  However, although our
sample is limited to only eleven galaxies, the kinematics and spatial
distribution of the stars in the radio detected sub-sample do not
stand out from the radio-undetected systems, suggesting that AGN
feedback in E+A galaxies does not play a dramatic role in defining
their properties, and/or that its effects are short.

Whichever process was responsible for the E+A signatures rotational
motion is seen in the majority of our sample, and their dynamics,
stellar masses and luminosities are consistent with those expected for
the progenitor population of (at least a subset of) representative,
early type galaxies.  Overall, these observations provide new and
detailed constraints on the kinematics and spatial distribution of the
A-stars and gas in E+A galaxies, and hence constraints on galaxy
formation models which aim to test the variety of physical mechanisms
which trace the evolutionary sequence linking gas-rich, star-forming
galaxies to quiescent spheroids.

\section*{acknowledgments}

We gratefully acknowledge the referee for their constructive report
which significantly improved the content and clarity of this paper.
AMS acknowledges an STFC Advanced Fellowship.  AIZ acknowledges funding
from NASA ADAP grant NNX10AE88G and also thanks the Institute of
Astronomy at Cambridge University and the Center for Cosmology and
Particle Physics at New York University for their hospitality during
the completion of this paper.  We thank Dave Alexander, Alastair Edge
and Russel Smith for numerous useful conversations.  These observations
are based on programs GS-2005B-Q-15, GN-2005B-Q-17, GN-2006B-Q-46,
GN-2007A-Q-27 \& GS-2007A-Q-22 from observations obtained at the Gemini
Observatory, which is operated by the Association of Universities for
Research in Astronomy, Inc., under a cooperative agreement with the NSF
on behalf of the Gemini partnership: the NSF (US), STFC (UK), the NRC
(Canada), CONICYT (Chile), the ARC (Australia), CNPq (Brazil) and
CONICET (Argentina).

\bibliographystyle{apj}
\bibliography{/Users/ams/Projects/ref}
\bsp

\end{document}